\documentclass[a4paper,12pt]{article}
\usepackage[utf8]{inputenc}
\usepackage{tikz}
\usepackage{tikz-3dplot, tkz-euclide}
\usetkzobj{all}
\usetikzlibrary{decorations.pathreplacing,shapes,calc,positioning}
\tdplotsetmaincoords{60}{60}
\usepackage{amsmath}
\usepackage{graphicx,subcaption}
\usepackage{amssymb}
\usepackage{adjustbox}
\usepackage{multirow}
\usepackage{color}
\usepackage{float, indentfirst}
\usepackage{soul}
\usepackage{booktabs}
\usepackage{natbib}
\usepackage{upgreek}
\usepackage{bm}
\usepackage{url}
\usepackage{mathrsfs,mathtools, esvect}
\usepackage{geometry}

\newcommand{\danilo}[1]{{\color{olive} Danilo: #1}}

\date{}

\title{Wisdom of the crowds forecasting the 2018 FIFA Men's World Cup}
\author{Marco Inácio, Rafael Izbicki, Danilo Lopes,\\ Luis Ernesto Salasar, João Poloniato \\ and Marcio Alves Diniz}

\begin{document}

\maketitle

\begin{abstract} The FIFA Men's World Cup  Tournament (WCT) is the most important football (soccer) competition, attracting worldwide attention. 
A popular practice among football fans in Brazil is to organize contests in which each participant informs guesses on the final score of each match.
The participants are then ranked according to some scoring rule. 
Inspired by these contests, we created a website to hold an online contest, in which participants were asked for their probabilities on the outcomes of upcoming matches of the WCT. After each round of the tournament, the ranking of all users based on a proper scoring rule were published. This paper studies the performance of some methods intended to extract the {\it wisdom of the crowds}, which are aggregated forecasts that uses some or all of the forecasts available. The later methods are compared to simpler forecasting strategies as well as to statistical prediction models. Our findings corroborate the hypothesis that, at least for sporting events, the {\it wisdom of the crowds} offers a competitive forecasting strategy. Specifically, some of these strategies were able to achieve high scores in our contest.
\end{abstract}

\section{Introduction}

Since its early days professional sports have been object of betting and gambling on the final outcome of their events. 
Although we may imagine Roman citizens betting on the result of a gladiator match, it is not hard to find registers of an organized betting industry devoted to sporting (and other kinds\footnote{\citep{rhode2013} describes the history of political betting markets.} of) events in the nineteenth century in Europe and U.S.A.  

In the case of association football---or soccer---the first pools, firms specialized in betting, appeared in the 1920's in England. After World War II, forecasting contests on football matches---usually sponsored by the state---spread in other European countries such as Italy (Totocalcio, 1946) and Spain (La Quiniela, 1948). 
In Brazil the first official contest related to football was created only in 1970 (Loteria Esportiva), but similar contests were created all over Latin America.

The most popular system was the $1\times 2$ bet, in which the bettor has to choose the final outcome of the match: ``$1$'' if she believes the home team will win, ``$2$'' if she believes the visiting team will win and ``$\times$'' if the she believes the match will end tied.
Usually the bettor that rightly picks the larger number of matches wins the contest, sharing the pool with others that picked the same number of correct guesses.

The popularity of such contests and their relation with probability assessments led the Italian statistician Bruno de Finetti ($1906-1985$) to idealize a similar contest, in which bettors would inform their probabilities for each possible result: victory of the home team, draw or victory of the visiting team.\footnote{De Finetti implemented such a kind of contest among a group of students and faculty members of the University of Rome in the 1960's. Other contests on probabilistic forecasts related to results of American football were implemented in the 1960's, but only for academic research. They are reported in  \cite{winkler1971probabilistic}.}
In 2017, the website FiveThirtyEight promoted exactly the contest idealized by de Finetti, but considering only matches of American football.

Inspired by these initiatives, we promoted an online contest where participants or forecasters\footnote{We also refer to them as ``users'' (of our website).} would inform their probabilities on the matches of the 2018 Men's World Cup Tournament (WCT), played in Russia between June 14 and July 15.
To give more incentive to participants, we informed that they would compete against two ``mathematical models'' whose forecasts were publicly available on websites.\footnote{Previsão Esportiva (\url{www.previsaoesportiva.com.br}) and Chance de Gol (\url{http://www.chancedegol.com.br/}).}
The forecasts were scored according to a proper scoring rule (a linear function of the Brier score) in order to be ranked.

The goal of this paper is to use the data collected in the contest to investigate the performance of different methods of aggregating the forecasts made by the participants. In particular, we wish to test if the aggregation strategies can perform better than statistical models that make forecasts based on the score of previous matches.

The fact that combining forecasts often leads to good results has been named \emph{Wisdom of the Crowds} (WOC) \citep{budescu2014identifying, davis2014crowd, Surowiecki2014}, which  has been applied to several fields including cosmology \citep{freeman2013new,lintott2008galaxy}, medicine \citep{raykar2010learning}, natural language processing  \citep{snow2008cheap}, and computer vision \citep{welinder2010online}. See \citep{Surowiecki2014} and references therein for other examples.
Although it is common to aggregate forecasts by unweighted averages \citep{esteves2017teaching, makridakis1983averages}, many other methods are based on assigning different weights to each forecast \cite{genest1986}. These weights can be computed by evaluating how much opinions from different forecasters differ among each other \citep{budescu2014identifying,dawid1979maximum}, or by evaluating the past performance of each forecaster \citep{vovk2009, cesa2006prediction}. Other approaches take into account additional information about each forecaster that can be correlated to their performance \citep{izbicki2013learning, raykar2010learning, yan2012modeling}. See \citep{frenay2013classification,olsson2015comparison} and references therein for a review of some approaches.

The remaining of the paper is organized as follows.
Section \ref{sec:methodology} presents all methods connsidered, briefly describing the scoring rule use, the statistical models mentioned above and the aggregation methods used to average the forecasts.
Section \ref{sec:results} reports the results of the contest and evaluates the performance of the methods. 
It also presents simulations that evaluate how robust was the final ranking of the contest. Section \ref{sec:conclusions} concludes the paper.
Technical details and more information about the contest are in the appendices.


\section{Methods}
\label{sec:methodology}

The participants of the contest had to access the website \url{fifaexperts.com} where, after registering, they were able to inform their probabilistic forecasts for the results of all scheduled matches of the WCT.
More specifically, they had to inform, for each match, a vector $P = (P_1,P_2,P_3)$, where $P_1$ denotes the probability of victory of the first team, $P_2$ the probability of victory of the second team and $P_3$ the probability of a draw.
The website enforced the constraints $P_1+P_2+P_3=1$ and $P_i\geq 0$, $i=1,2,3$.\footnote{See Appendix \ref{sec:website} for more information about the website and on how the participants informed their forecasts.}

After the end of each match the reported forecasts were numerically ranked by a scoring rule, which quantitatively measures how ``far'' the forecast was from the match outcome. We adopted as our scoring rule the Brier or quadratic score \cite{Brier1950},\footnote{As reported by \cite{machete2013}, there are several different metrics one can use to rank probabilistic forecasts.} which is the squared Euclidean distance between the forecast and the outcome of the event, the football match in our case. Applying a suitable linear transformation to the Brier score we obtain a standardized score between 0 (the worst possible forecast) and 100 (the best score, i.e. the one associated with a forecast that assigns probability one to the observed outcome of the match).\footnote{See Appendix \ref{sec:score} for the mathematical definition and properties of the Brier score.}

As mentioned above, we announced that two of the participants would be statistical or probabilistic models, namely

\begin{itemize}
\item \textbf{Chance de Gol.}  The statistical model adopted by  \url{http://www.chancedegol.com.br/} assumes a bivariate Poisson regression model for the final score of matches considering offensive and defensive factors of each team; see  \citep{diniz2019comparing} for details; and
\item \textbf{Previsão Esportiva.}  The statistical model adopted by \url{www.previsaoesportiva.com.br} is similar to that of ``Chance de Gol", but includes expert information in the analysis; see \citep{lee1997modeling, diniz2019comparing} for details.
\end{itemize}

During the WCT we were informed that at least other two participants reported forecasts of statistical models:

\begin{itemize}
\item \textbf{Esportes em números.} The statistical model used by \citep{fgv} (\url{http://www.fgv.br/emap/copa-2018/}), which is based on the model proposed by \citep{maher1982modelling}
and estimates
 the inherent offensive and defensive strengths of each team in a Poisson model; and
\item \textbf{Groll et al.} \citep{groll2018prediction}, which is based on random forests using covariates such as economic and sportive factors of each country.

\item We have also included as participant the forecasts provided by the website  \textbf{FiveThirtyEight}.
\end{itemize}

We will compare
the forecasts of all the participants (including the statistical models mentioned above) with the following aggregation strategies:

\begin{itemize}
\item \textbf{Top-$n$}.
Arithmetic average of the forecasts made by the top-$n$ participants (i.e., the $n$ participants with best score) before a specific match. We considered $n=1, 5, 10, 20$.
\item \textbf{Local wisdom}.
Average of all forecasts submitted for a given match.
\item \textbf{Global wisdom}.
Betting odds were collected from  18 online betting websites and the respective outcome probabilities were calculated using basic normalization \citep{vstrumbelj2014determining}. 
The forecasts of the best three\footnote{According to their scores up to that point.} websites were then averaged and reported as one forecast.
\item \textbf{Budescu and Chen}. The
approach proposed by \citep{budescu2014identifying}.
In our context, in the $(N+1)$-th match, this approach assigns to each forecaster $j$ a contribution factor of $C_j:=\sum_{i=1}^N (S_i-S_i^{-j})/N$, where $S_i$ is the score of the above-mentioned local wisdom strategy for the $i$-th match, and  $S_i^{-j}$ is the score of the local wisdom strategy for the same match \emph{removing forecaster $j$}. The aggregated forecast is given by the weighted average of the forecasts with positive $C_j$'s, using the latter constants as the weights.
\item \textbf{ISP-}$\eta$.
This is an \emph{individual sequence prediction} (ISP) approach proposed by 
\citep{cesa2006prediction}.
This approach is based on a weighted average of the forecasts given by each user. More precisely, the forecast for match $t$ is
    \begin{align*}
\widehat{p}_t :=\frac{\sum_{j=1}^k w_{j,t}f_{j,t}}{\sum_{j=1}^k w_{j,t}},
\end{align*}
where $w_{j,t}\geq 0$ is the weight given to user $j$ for that match. 
The weights are taken to be $w_{j,t}\propto \exp(\eta R_{j,t-1})$
    where $R_{j,t-1}$
    is the regret
    for the $j$-th user up to match $t-1$, defined as
$$R_{j,t-1}=\sum_{t_0=1}^{t-1} \left[ l(\widehat{p}_{t_0},y_{t_0})- l(f_{j,{t_0}},y_{t_0})\right],$$
where 
$f_{j,t_0}$ is the forecast of the $j$-th user for match ${t_0}$ and $l$ is a loss function (in our case, the negative value of the score defined in Equation \ref{eq:our_score}).
$\eta$ must be chosen by the user;
we have considered four values 0.001, 0.01, 0.1 and 1.
Under some conditions, this method corresponds to the multiparty Bayesian approach \citep{cesa2006prediction,pulgrossi2017comparison}.
\end{itemize}

As a baseline for comparisons, we have also included the following simple strategies:
\begin{itemize}
    \item \textbf{Monkey.} It randomly chooses a point on the simplex,\footnote{Imagine a monkey throwing darts at the 2-simplex.}
that is, the forecast is a uniformly distributed vector over the $2$-simplex: a Dirichlet distribution with parameter vector $(1,1,1)$.
\item \textbf{Edges.} 
It randomly chooses a point at one of the edges of the simplex, meaning that the probability of one of the results (victory of team 1, victory of team 2 or draw) is set to 0 and the other probabilities are randomly drawn from the remaining possible values.

\item  \textbf{Vertices.} 
It randomly picks one of the vertices $(1,0,0)$, $(0,1,0)$ and $(0,0,1)$ of the simplex, i.e., randomly selects a forecast that gives total certainty to one of the possible outcomes.
\item   \textbf{Maxi-min.} 
    It assigns equal probabilities for every possible result (i.e., the forecast is $(1/3,1/3,1/3)$ for all the matches).
This   is the non-randomized maximin strategy; see  Appendix \ref{sec:maximin}  for a proof.
\end{itemize}

\section{Results}
\label{sec:results}

\subsection{Descriptive Analysis}

At the end of the WCT, there were $511$ registered participants in the contest, though not all submitted forecasts for the 64 matches.
Table \ref{tab:sum} summarizes the number of participants according to the number of submitted forecasts, recalling that the group stage had $48$ matches, the round of $16$ had eight, the quarter-finals had four, the semi-finals two and the finals, two.

\

\begin{table}[!h]
\centering
\begin{tabular}{cc}
\hline
Number of forecasts & Number of partic. \\
\hline
$64$ & $57$  \\
$\geq 60$ & $104$ \\
$\geq 56$ & $139$ \\ 
$\geq 48$ & $217$ \\
\hline
\end{tabular}
\caption{Number of participants by number of forecasts}
\label{tab:sum}
\end{table}

The opening match between Russia and Saudi Arabia was the game with more submitted forecasts ($363$) and the match with fewer forecasts was the small final played by Belgium and England ($101$ forecasts).
The number of submitted forecasts for each matchup is displayed in Figure \ref{fig:forecasts} and can be seen in Tables \ref{tab:group} and \ref{tab:knock} in Appendix \ref{sec:extra_tables}.
Figure \ref{fig:forecasts} clearly shows that, after the group phase, several participants did not submit forecasts for the final matches.
We believe this happened because some lost interest due to their poor performance in the group phase. Also, participants had the option of submitting forecasts for the entire group phase before the WCT started, while in other phases the participants had to submit forecasts regularly after a batch of upcoming matches had been decided.

\begin{figure}[!h]
    \centering
    \includegraphics[scale=0.6]{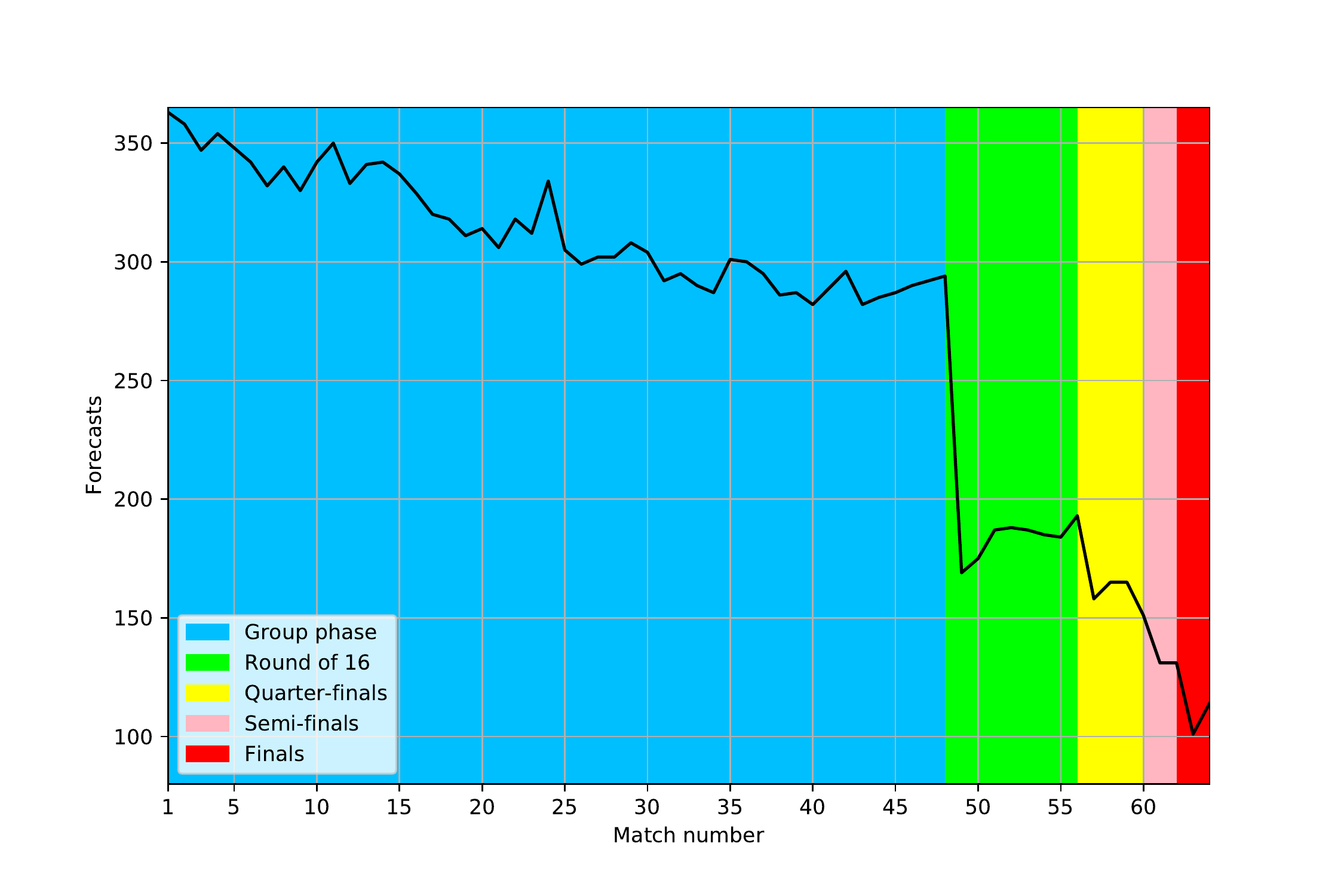}
    \caption{Number of forecasts for each match}
    \label{fig:forecasts}
\end{figure}

\begin{figure}[h!]
\centering
\begin{subfigure}{0.35\linewidth}
\includegraphics[scale=0.23]{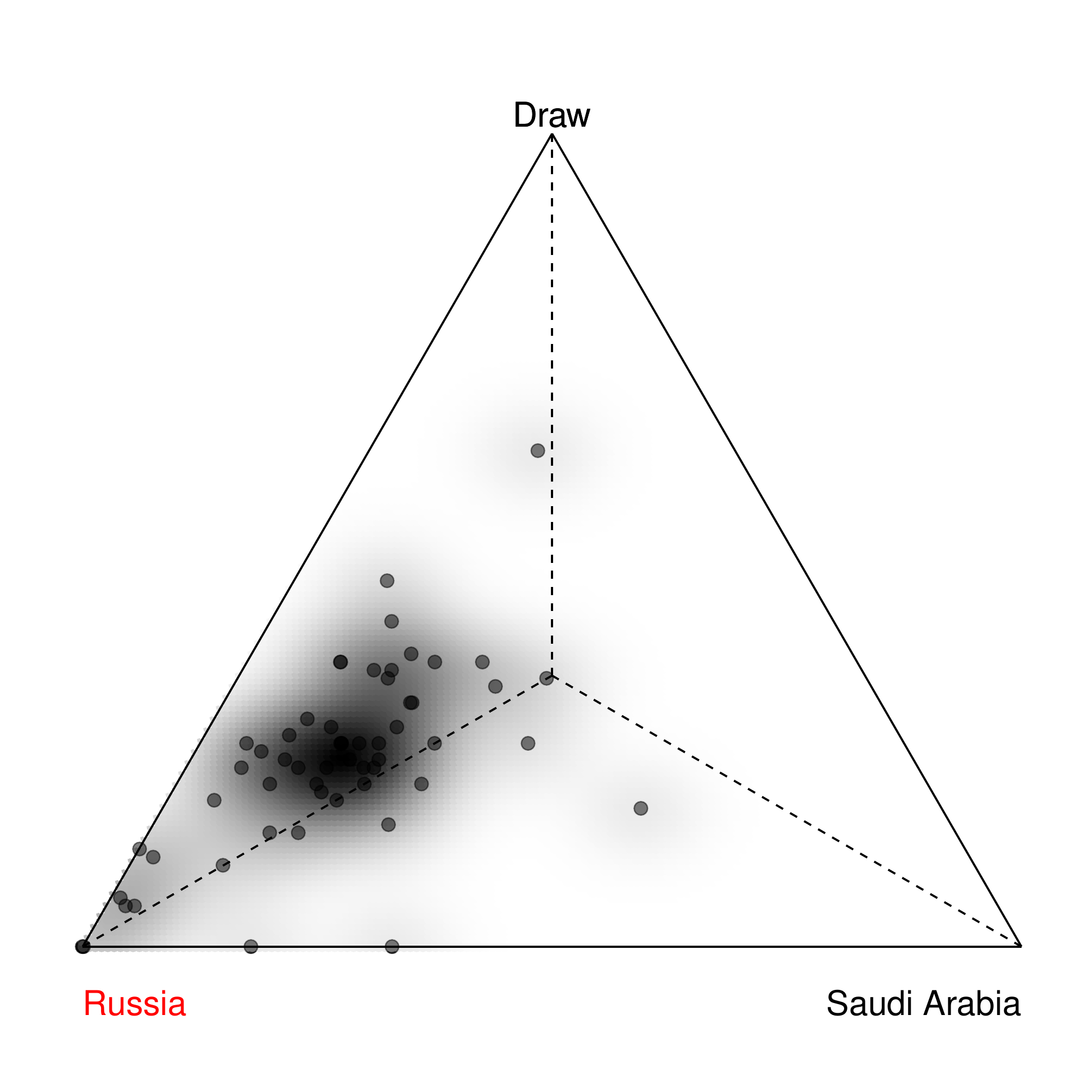}
\end{subfigure}
\begin{subfigure}{0.35\linewidth}
\includegraphics[scale=0.23]{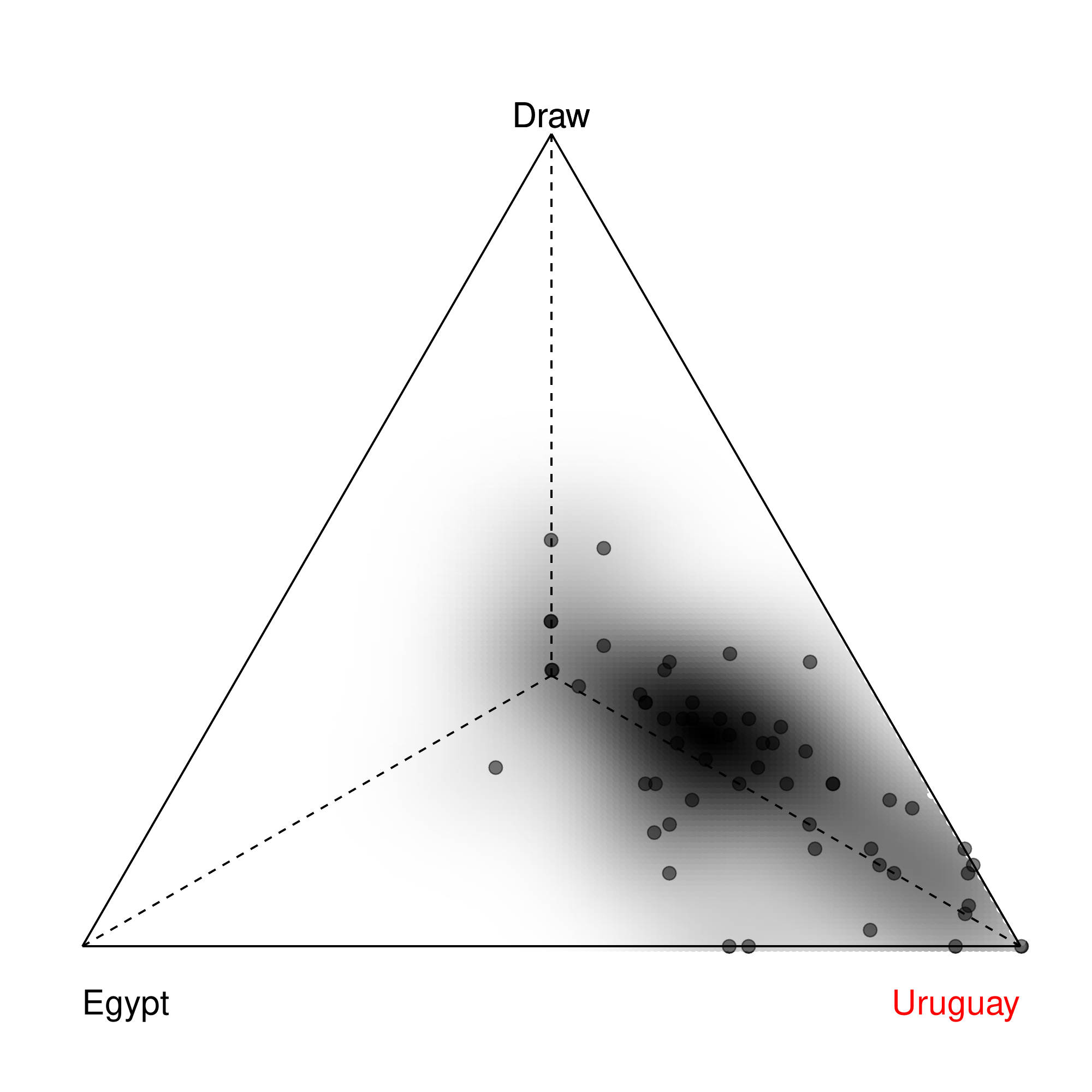}
\end{subfigure}
\begin{subfigure}{0.35\linewidth}
\includegraphics[scale=0.23]{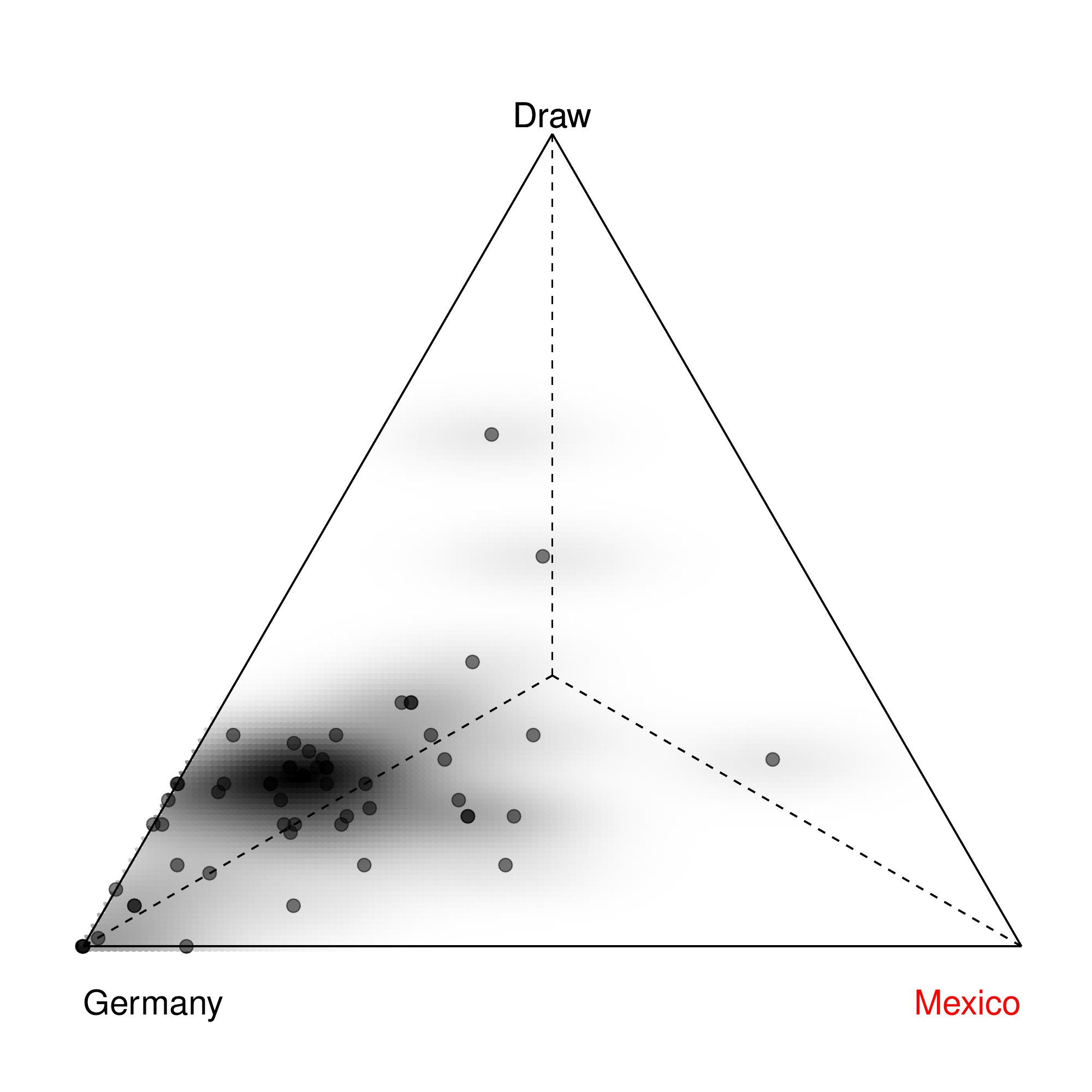}
\end{subfigure}
\begin{subfigure}{0.35\linewidth}
\includegraphics[scale=0.23]{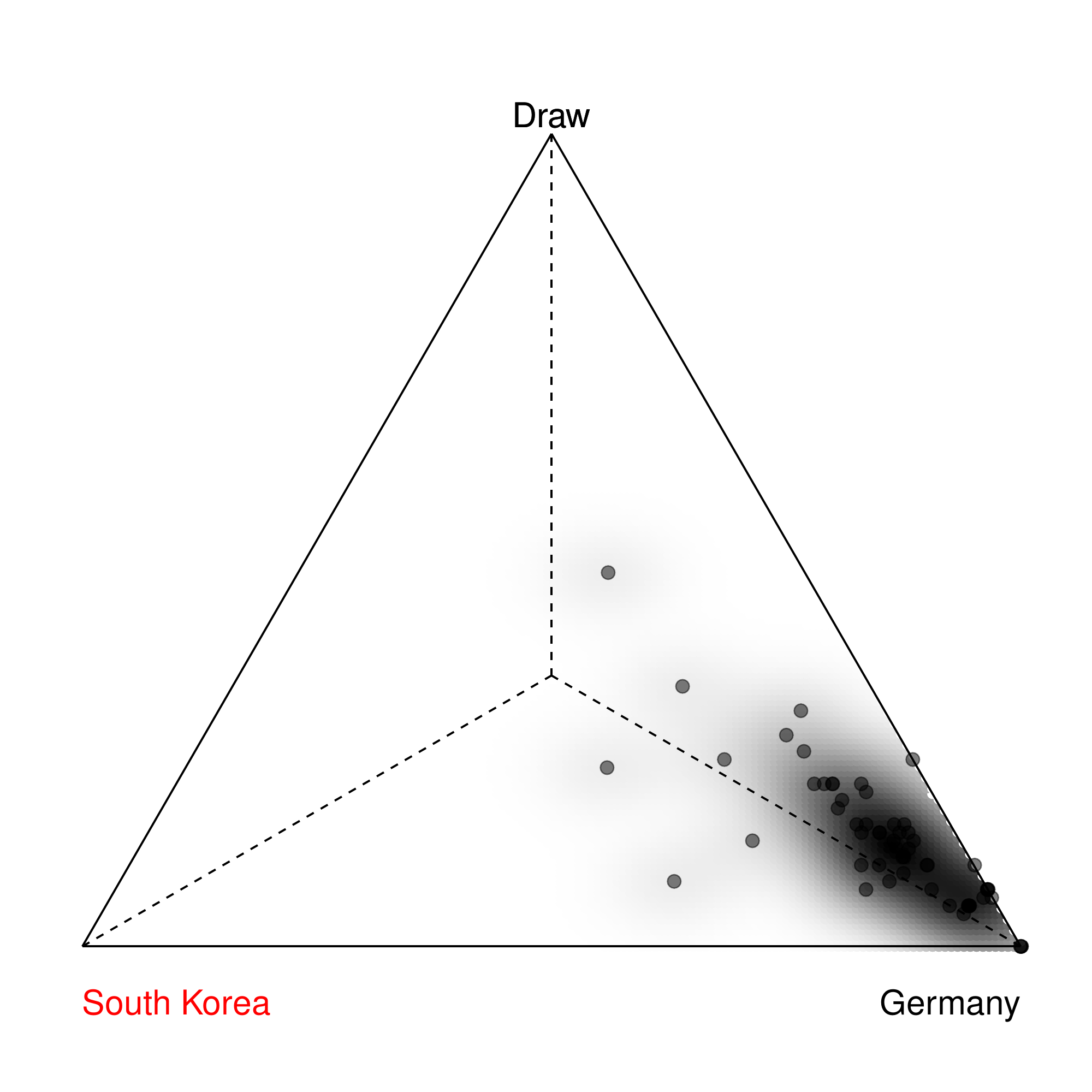}
\end{subfigure}
\caption{Forecasts made by each user for: Russia vs S. Arabia (top left), Egypt vs Uruguay (top right), Germany vs Mexico (bottom left) and S. Korea vs Germany (bottom right). Red vertex indicates the team that won the match.}

\label{fig:ternary}
\end{figure}

Figure \ref{fig:ternary} shows graphical illustrations of the forecasts submitted to four matches.
The top row displays 2 matches won by highly favourite teams, while the bottom row shows 2 matches won by underrated teams . The red vertex indicates the final result of the match and the intersection of the three dotted lines shows the maxi-min forecast $(1/3,1/3,1/3)$.
The top row displays the forecasts for Russia versus Saudi Arabia (left) and Egypt versus Uruguay (right).
The bottom row shows the forecasts for Germany versus Mexico (left) and South Korea versus Germany (right).

\begin{figure}[h!]
\centering
\begin{subfigure}[All users]{0.45\linewidth}
\includegraphics[scale=0.3]{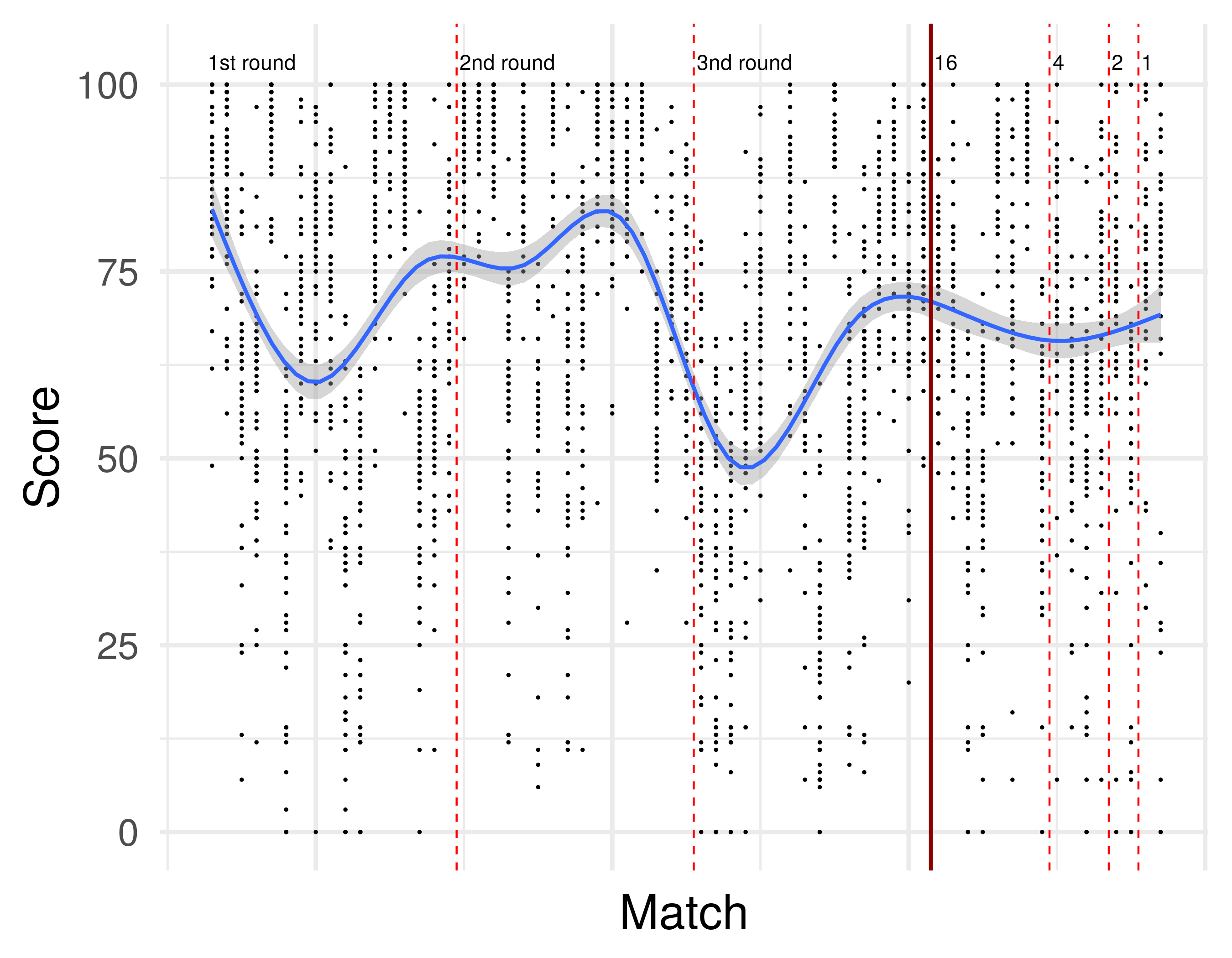}
\caption{All users}
\end{subfigure}
\begin{subfigure}[Top users]{0.45\linewidth}
\includegraphics[scale=0.3]{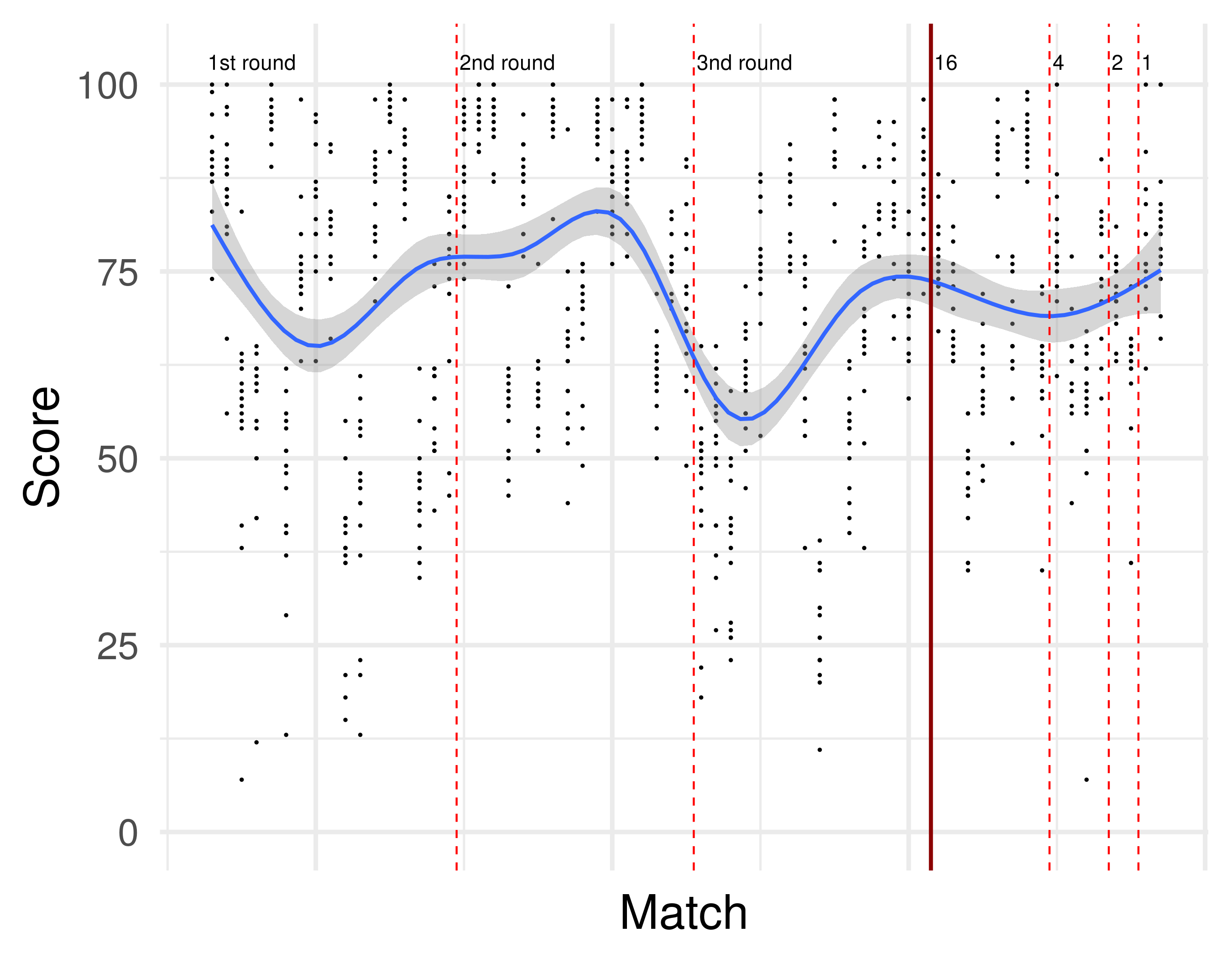}
\caption{Top users}
\end{subfigure}
\caption{Score that each user achieved for each match. Left panel contains all users that participated in the contest; right panel contains only the top 15.}
\label{fig:performance_by_time}
\end{figure}

Figure \ref{fig:performance_by_time} shows the scores of each forecaster for each match, numbered from 1 to 64 as displayed on Appendix \ref{sec:extra_tables}.
The vertical red lines represent the different stages of the WTC and the blue line is the fit of a smooth spline.
The left panel shows the results for all users that participated in the whole contest and the right panel shows the results only for the best 15 forecasters. 
Both graphs indicate that the third round of the group stage was the hardest one to predict. 
This probably happened because in this round some teams may not play competitively depending on whether they are already qualified, and draws occur more often as they may qualify the two opposing teams in a match, which are factors that cannot be easily described in a forecasting model. 
This unpredictability of the the final round of the group stage is a well-known fact for football fans and sports media, which led \cite{Chater2018} to suggest changing the current format of the WCT.
 
We have also computed the assertiveness of the forecasts, i.e. how bold was a given forecast.
Our interest in this measure came from the fact that the Brier score, used in our contest to rank the forecasts, discourage very assertive beliefs. 
We investigated this effect by comparing  best-performing with worst-performing
users. In order to do so, we define the assertiveness of a forecast
$P\in\Delta_2=\{(p_1,p_2,p_3):p_1+p_2+p_3=1, \ p_1, p_2, p_3 \geq 0\}$ as 

\begin{align*}
    \frac{3}{2}\sum_{i=1}^3 \left(P_i-\frac{1}{3}\right)^2,
\end{align*}
a quantity that has values ranging from zero to $1$\footnote{The assertiveness of the maxi-min forecast $P = (1/3, 1/3, 1/3)$ is zero and the assertiveness of a forecast at one of the vertices of the simplex is 1.} Thus, a forecast with large assertiveness is closer to one of the vertices of the simplex.

\begin{figure}[h!]
\centering
\includegraphics[scale=0.45]{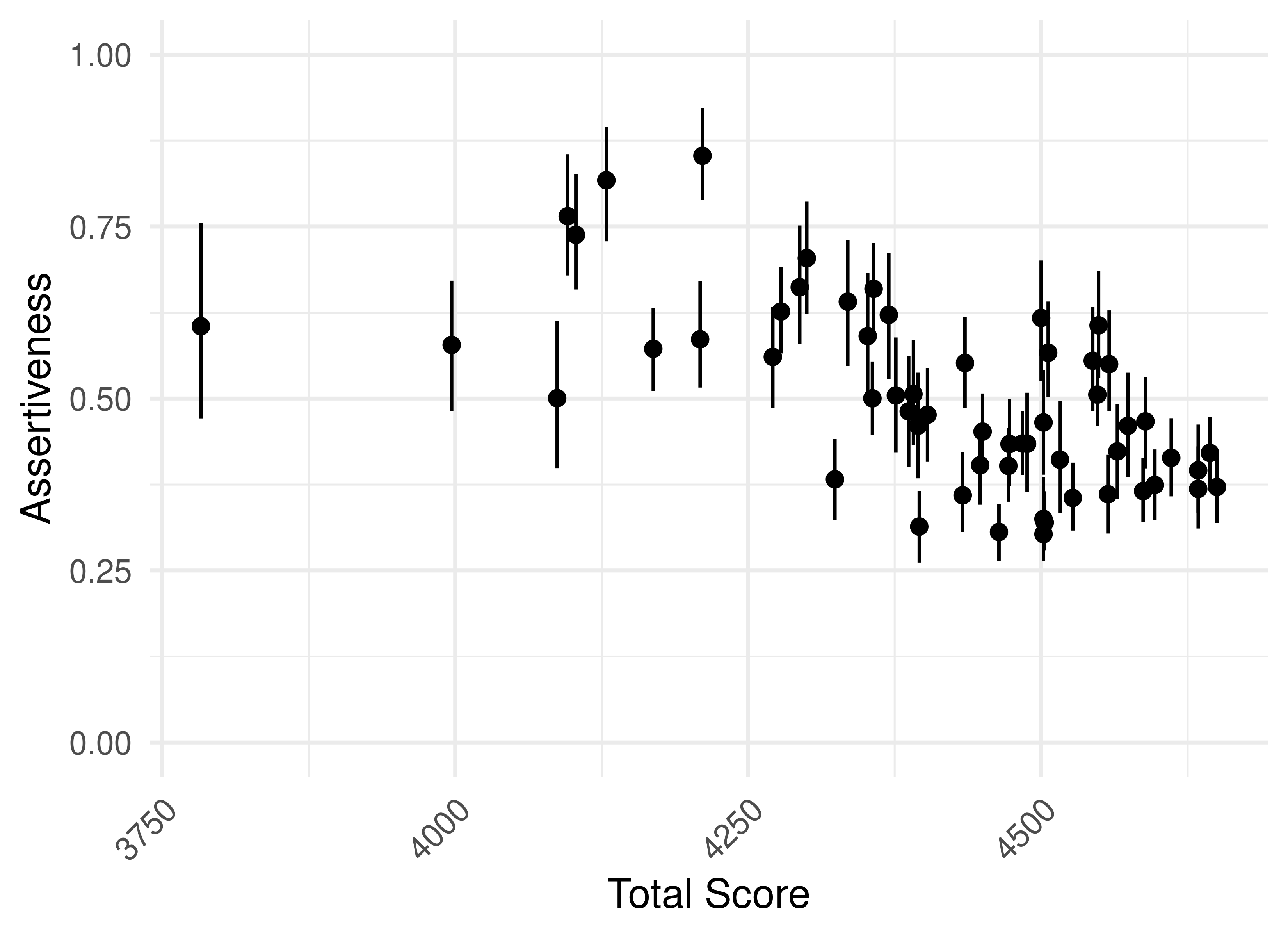}
\caption{Average assertiveness (and standard errors) of the forecasts made by each user as a function of their total score. }
\label{fig:assertiveness_by_total}
\end{figure}

Figure \ref{fig:assertiveness_by_total} shows how the average assertiveness of the forecasts made by each user as a function of the total score. In general, we see that users with the highest scores typically had less assertive forecasts.

\begin{figure}[h!]
\centering
\begin{subfigure}[All users]{0.45\linewidth}
\includegraphics[scale=0.3]{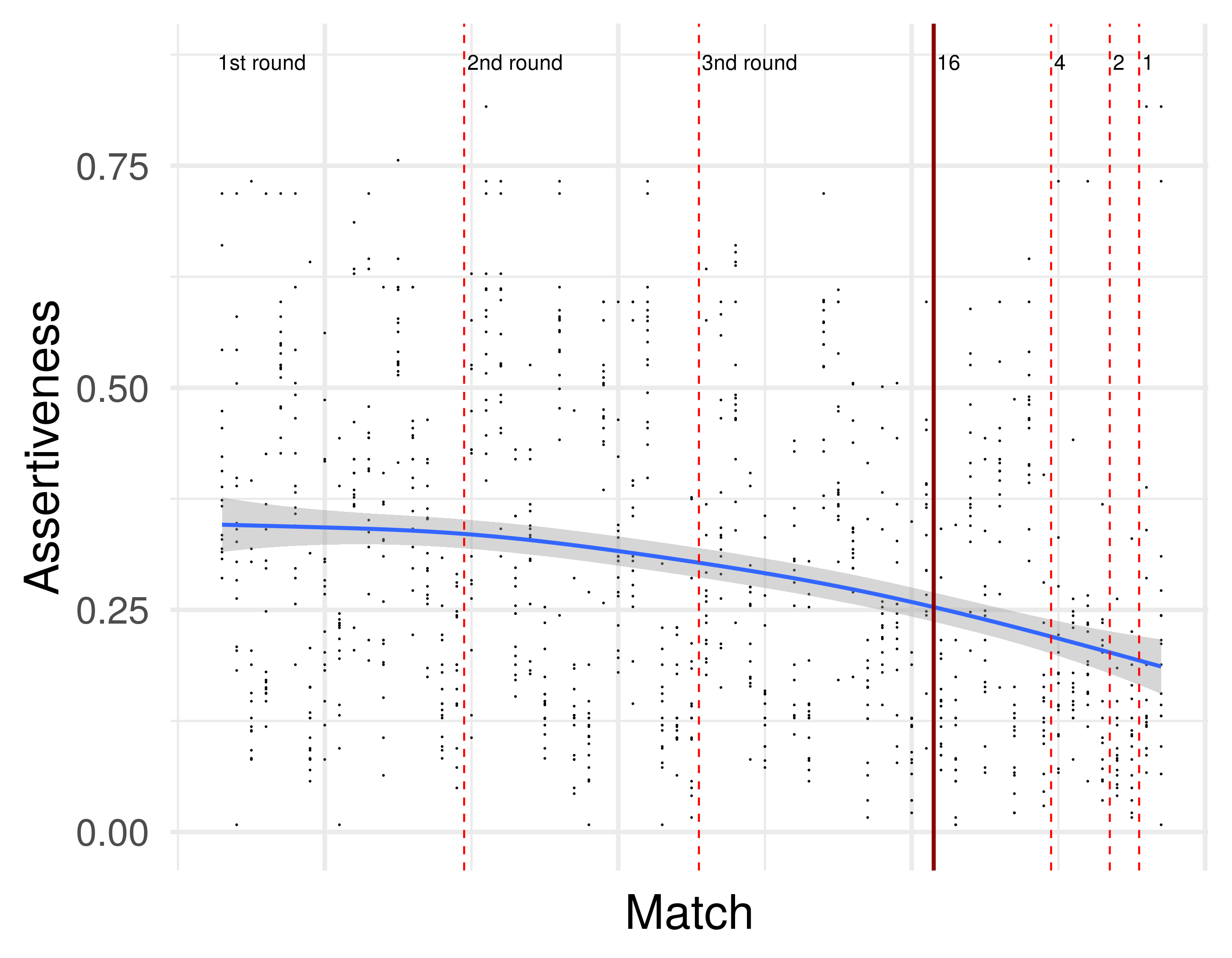}
\caption{Top users}
\end{subfigure}
\begin{subfigure}{0.45\linewidth}
\includegraphics[scale=0.3]{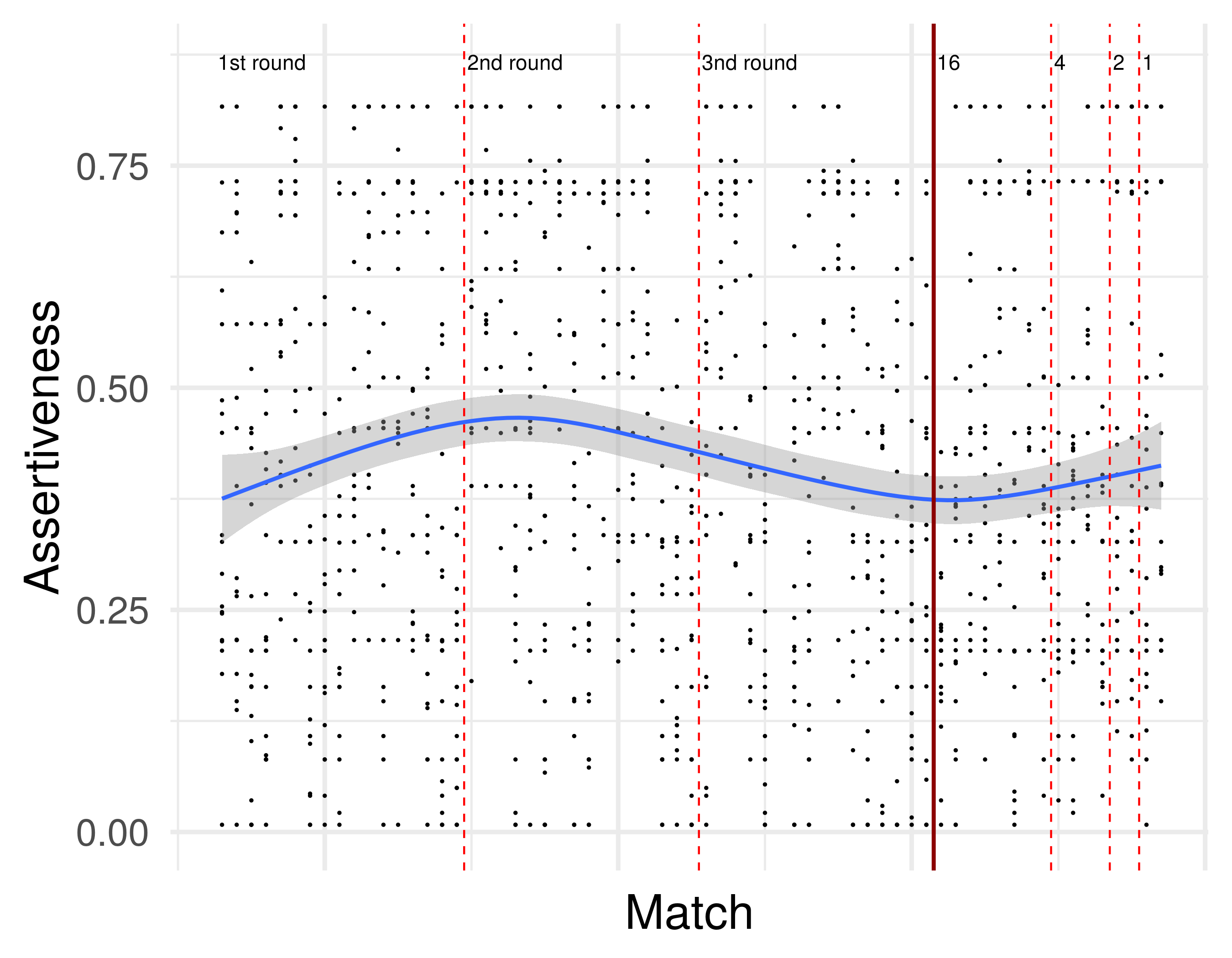}
\caption{Worst users}
\end{subfigure}
\caption{Assertiveness of the forecast made by each user for each match. Left panel contains 
only the top 15 users;
right panel contains only the worst 15.}
\label{fig:assertiveness_by_time}
\end{figure}

Figure \ref{fig:assertiveness_by_time} shows how the
assertiveness of each forecast varied according to the match.
While the left panel indicates that the assertiveness of the best forecasters decreased as the championship evolved  (i.e., the forecasts got closer to  (1/3,1/3,1/3)), the assertiveness of the worst forecasters was roughly stable. 
This was an expected feature, since the skill level of teams playing at later stages was similar and the best forecasters should know that before submitting their forecasts, while the worst users either were not aware of this or were acting as gamblers, trying to outscore the opponents to win the contest, and not informing their real beliefs.

\subsection{Aggregation  strategies}

In this section we evaluate how the aggregation strategies performed when compared to the participants.
Table \ref{tab:ranks} shows the ranking and final scores obtained for the WCT for all aggregation strategies and the participants described in Section \ref{sec:methodology}.
With exception of \textbf{Previsão esportiva}, statistical models had a good performance. In particular,
the top two scores were obtained by such models:
\textbf{Esportes em números} and \textbf{Groll et al.}. 

The best aggregation strategy was
\textbf{Global wisdom}, which is based on bets from external sources. In addition, the best aggregation strategy that only uses bets made on the website was
\textbf{Budescu and Chen}. The other aggregation strategies yielded poor results.
In particular, \textbf{Top-1}, which is very intuitive, 
had a poor performance: it was among the worst 50\% forecasters.
It is also interesting to note that the performance of
\textbf{ISP}  heavily depends on the tuning parameter $\eta$. 
For this application, taking
$\eta=0.01$ gave the best results. 
Still, the method was not among the top-10 best  forecasters.

\begin{figure}[hp!]
\centering
\includegraphics[scale=0.45]{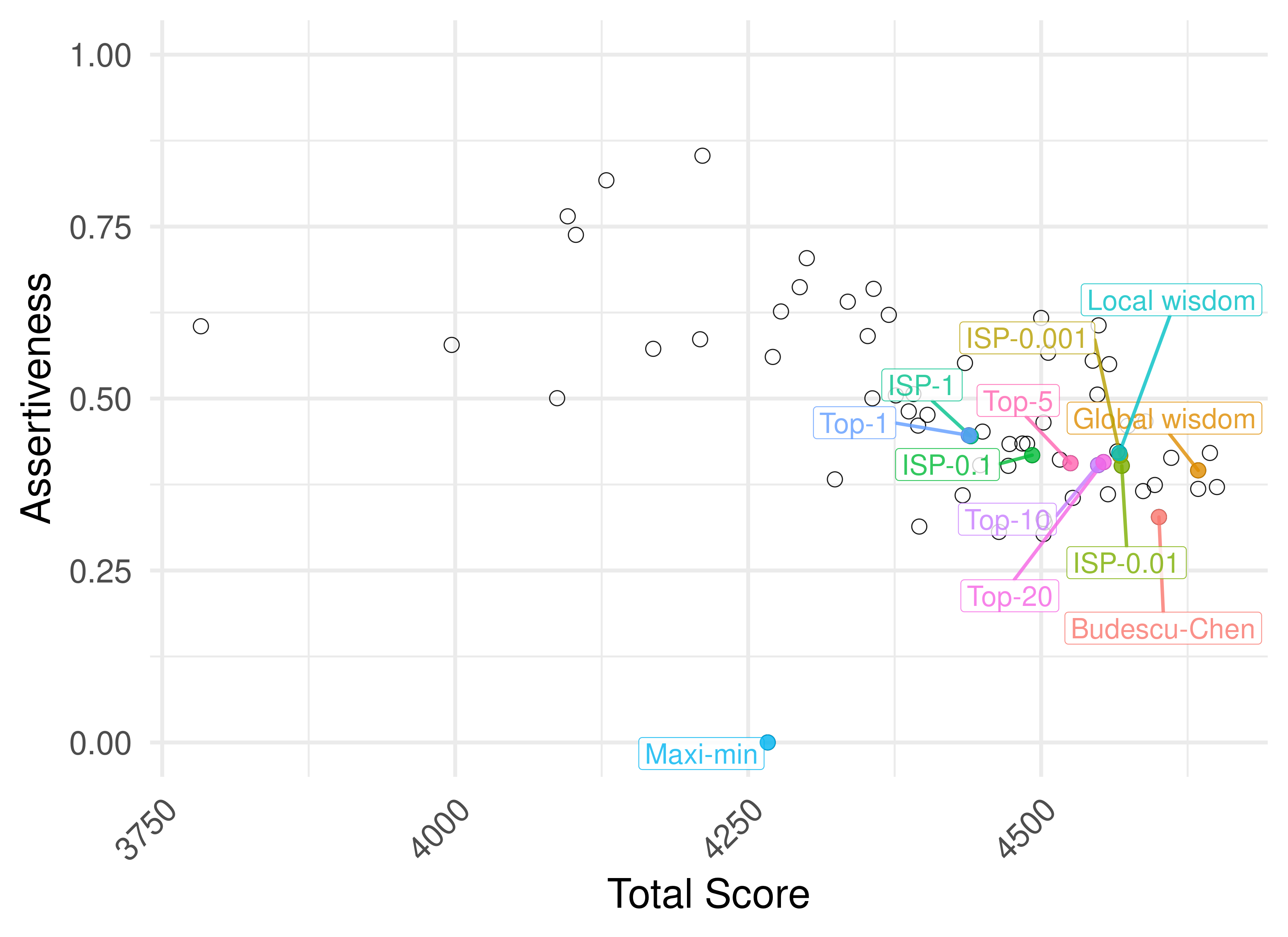}
\caption{Average assertiveness of the forecasts made by each participant and aggregation strategy as a function of their total score. }
\label{fig:assertiveness_by_total_identified}
\end{figure}

Figure \ref{fig:assertiveness_by_total_identified} shows the assertiveness and the overall score of each forecaster.
We identify the points associated to the aggregation strategies.  Apart from the maxi-min strategy, which is by definition the strategy with minimum assertiveness,
all the other aggregation strategies have similar assertiveness.

\begin{table}[hp!]
\centering
\caption{Performance of the aggregation strategies as well as some  participants.}
\begin{tabular}{lrrr}
  \hline
Forecaster & Total Score & Average score/game & Rank  \\ 
  \hline
Esportes em números & 4650 & 72.7 & 1.0 \\ 
  Groll et al. & 4644 & 72.6 & 2.0 \\ 
  Global wisdom & 4634 & 72.4 & 3.5 \\ 
  FiveThirtyEight & 4634 & 72.4 & 3.5 \\ 
  Chance de gol & 4611 & 72.0 & 5.0 \\ 
  Budescu and Chen & 4601 & 71.9 & 6.0 \\ 
  ISP-0.01 & 4569 & 71.4 & 11.0 \\ 
  ISP-0.001 & 4567 & 71.4 & 12.5 \\ 
  Local wisdom & 4567 & 71.4 & 12.5 \\ 
  Top-20 & 4553 & 71.1 & 17.0 \\ 
  Top-10 & 4549 & 71.1 & 18.5 \\ 
  Top-5 & 4525 & 70.7 & 23.0 \\ 
  ISP-0.1 & 4492 & 70.2 & 31.0 \\ 
  Previsão esportiva & 4450 & 69.5 & 37.0 \\ 
  ISP-1 & 4440 & 69.4 & 39.0 \\ 
  Top-1 & 4438 & 69.3 & 40.0 \\ 
  Maxi-min & 4267 & 66.7 & 59.0 \\ 
  Monkey & 3733 & 58.3 & 69.0 \\ 
  Edges & 3200 & 50.0 & 70.0 \\ 
  Vertices & 2133 & 33.3 & 71.0 \\ 
   \hline
\end{tabular}
\label{tab:ranks}
\end{table}

\subsection{Using simulation to evaluate user ranking}
\label{sec:user_simulations}


The total scores obtained by the
top performing methods on Table \ref{tab:ranks} are very close to each other, especially considering that they are  based only on 64 matches. This raises the question of whether it is possible to say that the  winner is indeed  the best forecaster, or  if was only a matter of luck.
In this section we evaluate this question using simulations. Here we only consider users; a similar analysis for aggregated strategies is done in Section \ref{sec:agg_simulations}.

In order to make our
simulations realistic, we first
used the probabilities assigned by the winner of the contest (\textbf{Esportes em números}) as if they were the true generating probabilities of the match outcomes. Using these probabilities, several independent tournaments are simulated and the rankings of each user are evaluated for each of these simulated tournaments. The idea is to check how many times the user that is the best one by definition (because it is used to generate the outcomes) is indeed the winner of the contest.

The details  are as follows: for each simulation, we generated the outcomes of the 64 matches independently according to the probabilities submitted by the actual winner of the contest; we then calculated the total score for each participant using these simulated outcomes and ranked the participants according to their simulated total scores; at the end of 100,000 simulations, we calculated the proportion of times a given participant ended up in a given position.





\begin{figure}[h!]
\centering
\includegraphics[scale=0.5]{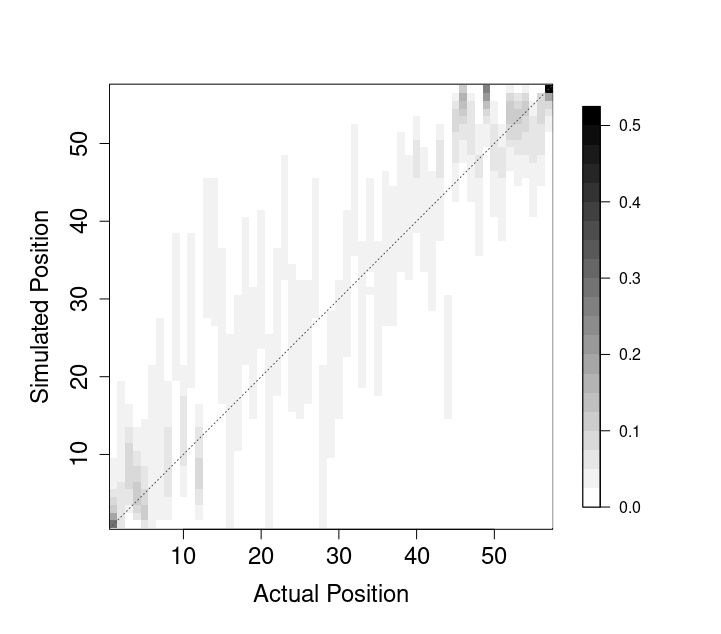}
\caption{Probability that a user who ended up in a given position (horizontal axis) during the contest with 64 forecasts would end up in some other position (vertical axis) if the forecasts given by the top user were true match probabilities.}
\label{fig:simulatedrankings}
\end{figure}

Figure \ref{fig:simulatedrankings} shows a heat map of estimated probabilities that a user with a given position (horizontal axis) in our observed contest would end up in a given position during simulations (vertical axis). The darker the pixel the higher the corresponding probability is. The dashed diagonal line is the equality line. As a reference, all the values of probability in the represented data matrix add up to one either by row or by column. 

The figure shows that
there is some association between the final position in the actual contest and the position of the same participant in the simulations (most of the shades in the graph are concentrated along the equality line).
However, this association is not very strong, as indicated by the lightness of those shades, with most of the probabilities around the equality line being below 10\%. Not only that, but the pixel in the lower left corner is not as dark as one would expect: the actual winner ended up in the first place only in 28\% of the simulations.
It is also worthwhile noting that the participant with the lowest final ranking has the darkest pixel in Figure \ref{fig:simulatedrankings} as a combination of her assertiveness and some distance of her forecasts to those of the other forecasters.
Table \ref{tab:simulatedranking} presents a summary of the results and corroborates our conclusion that the ranking of the contest does not clearly define who is indeed the best forecaster. 


\begin{table}[!h]
\caption{Percentages of simulations a given participant ended up in a certain position, and average and standard deviation of the positions in the simulations. The forecasts of 57 users for each one of the 64 matches in the WCT are fixed as the ones informed in the actual contest, but the match outcomes are simulated according to the forecasts of the actual winner \danilo{(\textbf{Esportes em números})}.}
\label{tab:simulatedranking}
\centering
\begin{tabular}{lccccccc}
\hline
& \multirow{2}{*}{1st} & \multirow{2}{*}{2nd} & \multirow{2}{*}{3rd} & \multirow{2}{*}{4th} & Other & Average  & SD of  \\
&  &  &  &  &  positions & Position & Position \\
\hline
Actual 1st place & 28.2\% & 17.7\% & 11.4\% & 8.2\%  & 34.5\% & 4.7 & 4.8\\
Actual 2st place & 5.1\%  & 6.1\%  & 6.1\%  & 5.8\%  & 76.9\% & 11.5 & 8.0\\
Actual 3rd place & 0.8\%  & 2.0\%  & 3.4\%  & 5.1\%  & 88.7\% & 9.5 & 4.2\\ 
Actual 4th place & 2.5\%  & 6.4\%  & 9.8\%  & 11.7\% & 69.6\% & 7.1 & 4.1\\
\hline
\end{tabular}
\end{table}

In order to check the robustness of this conclusion, we repeat the simulations using different users to generate the match outcomes. Now, for each of the top four participants of the actual contest, we ran 100,000 simulations using her forecasts as the true probabilities for the matches and then calculated the proportion of times this participant end up in a given position. 

Each row of Table \ref{tab:diffenttruths} corresponds to a different set of simulations and shows the distribution of the position of a given participant when her own forecasts were the true probabilities. 
The user who held second place in the actual contest (\textbf{Groll et al.}) had a similar pattern when she knew the truth compared to the actual winner (\textbf{Esportes em números}), with an even higher probability of taking the first place. On the other hand, the actual third and fourth places (\textbf{Global wisdom} and \textbf{FiveThirtyEight}, respectively) had higher probabilities of being in second, third or fourth places than being in the first place when each of them knew the truth. They all had the highest probabilities of ending up in each of the top positions among participants, but the distributions of their positions during simulations are concentrated on higher values, as indicated by their respective mean and standard deviation.

\begin{table}[!h]
\caption{Percentage of simulations a given participant ended up in a certain position, and average and standard deviation of the positions in the simulations, when her respective forecasts were true. The forecasts of 57 users for each one of the 64 matches in the WCT are fixed as the ones informed in the actual contest, but the match outcomes are simulated according to the forecasts of the participant in each row.}
\label{tab:diffenttruths}
	\centering
	\begin{tabular}{lccccccc}
		\hline
		& \multirow{2}{*}{1st} & \multirow{2}{*}{2nd} & \multirow{2}{*}{3rd} & \multirow{2}{*}{4th} & Other & Average  & SD of  \\
		&  &  &  &  &  positions & Position & Position \\
		\hline
		Actual 1st place & 28.2\% & 17.7\% & 11.4\% & 8.2\%  & 34.5\% & 4.7 & 4.8\\
Actual 2nd place & 37.4\% & 16.2\% & 9.7\% & 6.8\% & 29.9\% & 4.3 & 4.8\\
Actual 3rd place & 5.8\% & 10.4\% & 12.6\% & 12.9\% & 58.3\% & 5.8 & 3.5\\ 
Actual 4th place & 9.1\% & 13.0\% & 13.2\% & 12.2\% & 52.5\% & 5.6 & 3.9\\
\hline
\end{tabular}

\end{table}

The lack of stability in the rankings 
is in part due to the small number of matches. 
In order to test if with a larger number of matches 
a user who knows the truth would necessarily show her superiority,
we performed a bootstrap-like simulation:
we analyze the behavior of the forecasts in a situation with $n$ matches 
by taking a sample with replacement of size $n$ from the original 64 matches. 
Now, for a given number $n$ of matches and a given participant who knows the truth, we run 10,000 simulations such that, in each simulation, we sample $n$ matches with replacement, generate $n$ outcomes independently according to the truth (if the same match was selected more than once, it may have different outcomes during simulation), calculate the total score and rank all the participants according to their total score. 

\begin{figure}[h!]
\centering
\includegraphics[scale=0.6]{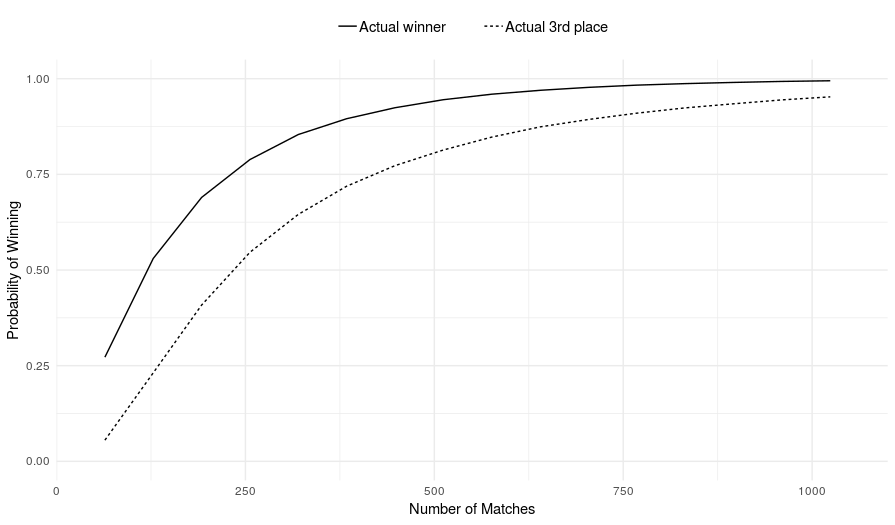}
\caption{Probability that the actual winner (solid line) and the actual third pĺace (dashed line) win the contest with when their respective forecasts were true versus the number of matches in the contest.}
\label{fig:morematches}
\end{figure}

Figure \ref{fig:morematches} shows the estimated probability of winning the contest (vertical axis) for the actual winner (\textbf{Esportes em números, }solid line) and the actual third place (\textbf{Global wisdom}, dashed line) when knowing the truth versus the number of matches (horizontal axis) during this simulated fictional contest. The actual third place was chosen here because of her very small chance of winning when knowing the truth with only 64 matches (5.8\%, as shown in Table \ref{tab:diffenttruths}). As expected, both lines increase and get closer to one as the number of matches increases. The actual first place needed a contest with approximately 576 matches in order to have a probability of 95\% of winning when knowing the truth, while the actual third place would need approximately 1,024 matches in order to have at least the same probability of 95\%.

Our simulations show that hundreds of matches would be necessary to tell two very good forecasters apart. Making this kind of comparison in real life, however, seems to be doable only for forecasting models, since human forecasters would find the task of providing hundreds of forecasts a very burdensome task.

 \subsection{Using simulation to evaluate aggregation strategies}
 \label{sec:agg_simulations}
 
Our findings show that WOC strategies were superior to most of the individual participants including the statistical models and the simpler strategies.
However one could ask, in the same spirit of the last subsection: how could one tell that these strategies were really the best ones and not just ``got lucky'' about the 64 matches of this particular tournament?
 
To answer this question, similarly to the procedure described in Subsection \ref{sec:user_simulations}, we simulated 100,000 fictitious tournaments using the forecasts of a given participant as the true generating mechanism for the match outcomes. Then, we compared the performance of the aggregation strategies {\bf Top-5}, {\bf Local Wisdom} and {\bf Budescu and Chen} separately to all the 57 participants (i.e., strategies are not compared among them when computing simulated positions).
 
The comparison was based on the proportion of simulations each strategy ended up in a given position under two different simulation scenarios for the 64 matches: (i) match outcomes are simulated from the forecasts of the actual winner (\textbf{Esportes em números}) and (ii) match outcomes are simulated from the forecasts of the actual third place (\textbf{Global wisdom}). Tables \ref{tab:simulatedWOC1} and \ref{tab:simulatedWOC2} present the results for the scenarios (i) and (ii), respectively. Observe that the proportions presented in Tables \ref{tab:simulatedWOC1} and \ref{tab:simulatedWOC2} are much smaller than those presented in Tables \ref{tab:simulatedranking} and \ref{tab:diffenttruths}, which is expected since it is difficult for the aggregation strategies to identify the good forecasters using only 64 matches. For both scenarios, \textbf{Top-5} is the riskier strategy, with higher chance of taking top positions, but also with higher variability, as indicated by the greater average and large standard deviation of the final position. \textbf{Local Wisdom} strategy is the most conservative one, with moderate values of average and standard deviation of final positions. 
 
\begin{table}[!h]
\caption{Percentages of simulations the selected aggregation strategies ended up in a given position relative to the 57 participants, including the respective average and standard deviation of the final positions. The match outcomes were simulated from the forecasts of the actual first place participant.}
\label{tab:simulatedWOC1} \centering
     \begin{tabular}{lccccccc}
     	\hline
    & \multirow{2}{*}{1st} & \multirow{2}{*}{2nd} & \multirow{2}{*}{3rd} & \multirow{2}{*}{4th} & Other & Average & SD of  \\
    &  &  &  &  &  positions & Position & Position \\
    \hline
     	Top-5 & 1.2\% & 1.9\% & 2.5\% & 2.7\%  & 91.7\% & 15.2 & 7.7\\
     	Local Wisdom & 0.2\%  & 0.9\%  & 2.1\%  & 3.8\%  & 93.0\% & 10.1 & 3.9\\
     	Budescu and Chen & 0.2\% & 0.8\% & 1.2\% & 1.6\%  & 96.2\% & 13.6 & 4.6\\
     	\hline
    \end{tabular}
\end{table}

\begin{table}[!h] \centering
\caption{Percentages of simulations the selected aggregation strategies ended up in a given position relative to the 57 participants, including the respective average and standard deviation of the final positions. The match outcomes were simulated from the forecasts of the actual third place participant.}
\label{tab:simulatedWOC2}
\begin{tabular}{lccccccc}
\hline
& \multirow{2}{*}{1st} & \multirow{2}{*}{2nd} & \multirow{2}{*}{3rd} & \multirow{2}{*}{4th} & Other & Average  & SD of  \\
&  &  &  &  &  positions & Position & Position \\
\hline
Top-5 & 2.1\% & 3.2\% & 3.7\% & 3.8\%  & 87.1\% & 13.6 & 7.9\\
Local Wisdom & 0.4\%  & 1.8\%  & 3.9\%  & 6.8\%  & 87.1\% & 8.3 & 3.5\\
Budescu and Chen & 0.5\% & 1.5\% & 2.3\% & 2.8\%  & 92.9\% & 12.5 & 5.1\\
\hline
\end{tabular}
\end{table}

In order to study the effect of increasing the number of forecasts in the contest on the performance of the three aggregation strategies, we created fictitious tournaments of $n$ matches by sampling with replacement from the 64 matches in the WCT and using their corresponding forecasts submitted to the contest, similarly to the procedure described in Subsection \ref{sec:user_simulations}. 

\begin{figure}[h!]
	\centering
	\includegraphics[scale=0.34]{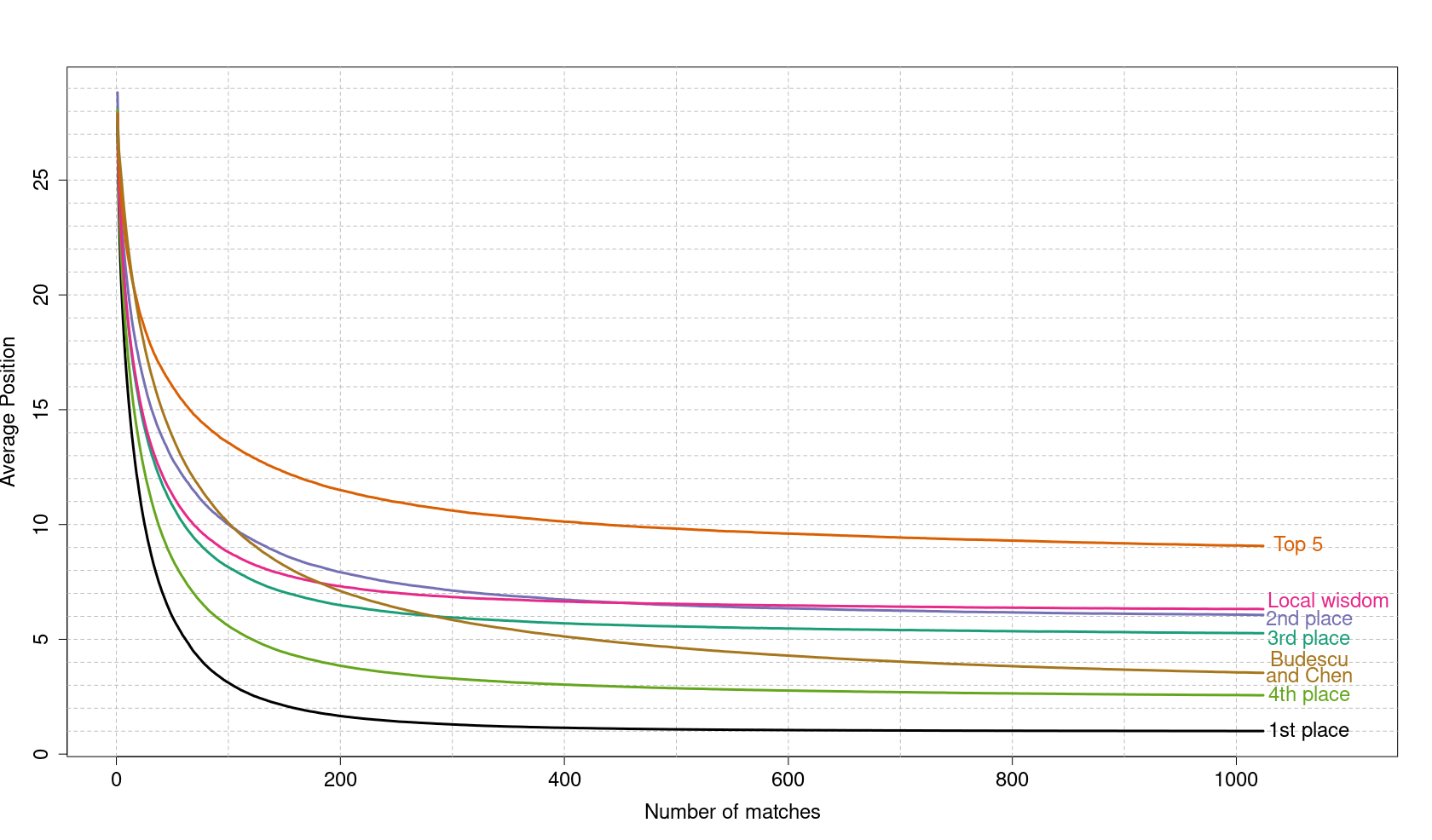}
	\caption{Average final position of a given participant (or strategy) in the simulations when forecasts of the actual winner were true versus the number of matches in the contest with 57 participants.}
	\label{fig:matchesWOC}
\end{figure}

For $n$ varying from 1 to 1,024, we ran 30,000 simulations of these fictitious tournaments assuming that forecasts of the actual winner were true and then computed the average final position for each one the actual top-4 participants as well as for our three WOC strategies (once again, we compare each strategy individually to the 57 users). Results are shown in Figure \ref{fig:matchesWOC}. All the strategies and top participants improve their average positions as we increase the number of forecasts in the contest, but the most noticeable improvement is certainly the one of the strategy \textbf{Budescu and Chen}. The intuitive reason for the good performance of this method is that its weighted average expression can correctly identify the good forecasters, as is the case of the actual winner, and at the same time disregard the misinformed participants, those with negative contribution factors. Learning about which participants bring relevant information and which does not is the main advantage of \textbf{Budescu and Chen} method, but it may take some hundreds of forecasts for its prevalence among strategies to appear.

\section{Final remarks}
\label{sec:conclusions}

The probabilistic previsions submitted to our website to forecast the matches of the 2018 Men's WTC revealed some interesting characteristics of the behaviour of such contests and of the performance of WOC strategies, considering different ways of opinion aggregation.

Our first finding shows that forecasters that were very assertive presented an overall performance that was not as good as the best forecasters.
As expected, the assertiveness of the best forecasters got closer to the naive prediction (1/3,1/3,1/3) as the WCT reached its final matches, since the teams that qualified to the final rounds were, theoretically, closely skilled.

Regarding the aggregation (WOC) strategies, the best ones (\textbf{Global wisdom} and \textbf{Budescu and Chen}) had an outstanding overall performance of the same level of the best forecasters, which adopted some sort of statistical model or algorithm.
However, this was not true for all of them, such as ISP-0.1, which ranked 31.
An important remark about the strategies \textbf{Global wisdom} and \textbf{Budescu and Chen} is that one should know all the submitted forecasts of an upcoming match to provide a forecast for the same match, which would not be feasible for a regular user of our website. 

Finally, our simulations revealed that a tournament with 64 matches, like the WTC, is not sufficient to identify the best forecasters since the observed performance of any given participant might be due to randomness. Thus, 
longer tournaments such as the English Premier League, which has 380 matches every season, may yield further insights on which aggregation strategies work best. However,
conducting an open contest of the same sort we did for the WTC would require incentive mechanisms to avoid individual participants of leaving the contest before its end.



\section*{Acknowledgments}

Marco Inácio is grateful for the financial support of CAPES: this study was financed in part by the Coordenação de Aperfeiçoamento de Pessoal de Nível Superior - Brasil (CAPES) -
Finance Code 001. Rafael Izbicki is grateful for the financial support of FAPESP (2019/11321-9)
and CNPq (306943/2017-4).

\bibliographystyle{plainnat}
\bibliography{main}

\newpage

\appendix

\section{The website}
\label{sec:website}

The easiest way to implement such a kind of contest is to allow forecasts to be informed online and automatically scored once the results of the matches are available.
Our website was heavily based on the page aired by FiveThirtyEight.com for the 2017 NFL season\footnote{The page for the 2018 season is https://projects.fivethirtyeight.com/2018-nfl-forecasting-game/.}.
The opening page had a short text explaining how the contest would work, and below the text there was a table with four tabs.

\begin{figure}[H]
\centering
\includegraphics[scale=0.25]{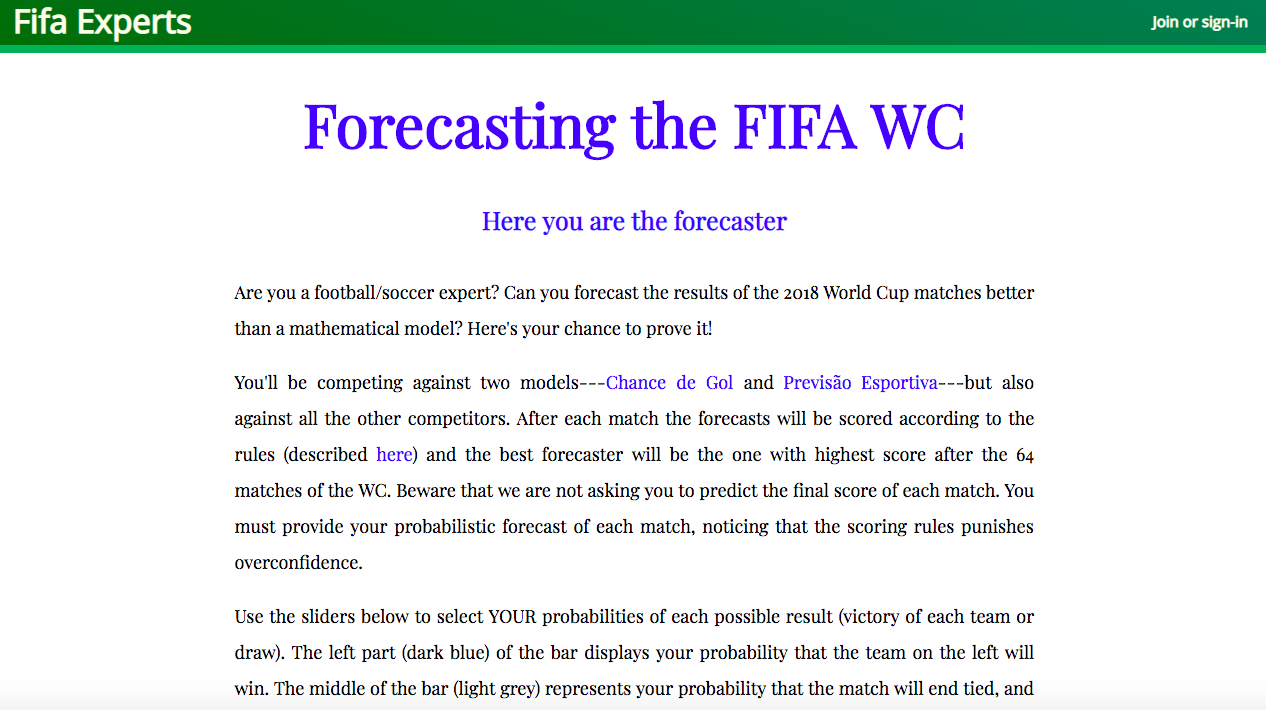}
\caption{Opening page}
\label{fig:1}
\end{figure}

The first tab was ``Submit your forecasts'', where all the scheduled matches were displayed and it was possible to inform the probabilities of each possible result moving two sliders on a bar.
The left part of the bar (in dark blue) displayed the bettor or forcaster's probability that the team on the left would win. The middle of the bar (in light grey) represented her probability that the match would end tied, and the right part of the bar (in dark grey) showed her probability that the team on the right would win.

\begin{figure}
     \centering
     \begin{subfigure}[b]{0.49\textwidth}
        \centering
        \includegraphics[width=\textwidth]{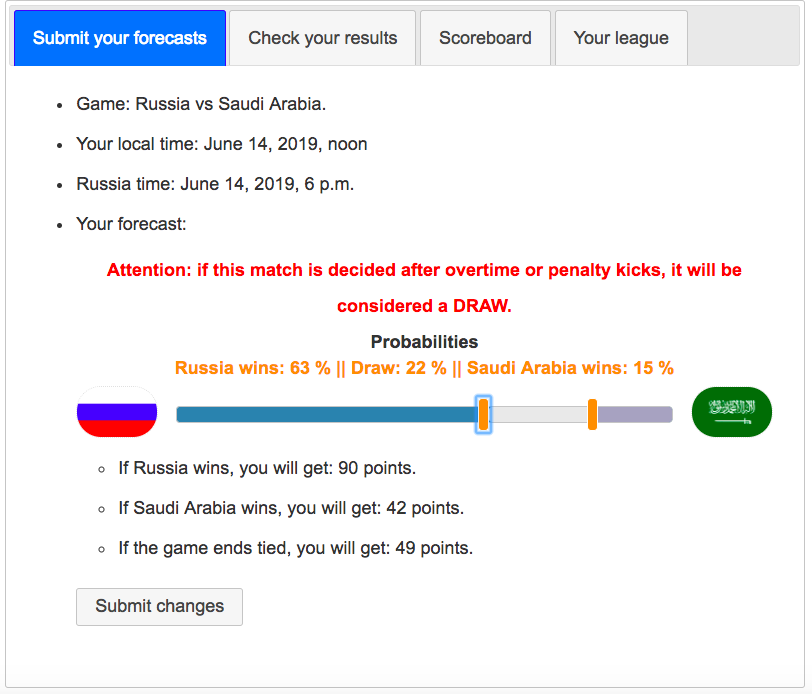}
        \caption{Entering probabilities}
        \label{fig:2}
     \end{subfigure}
     \hfill
     \begin{subfigure}[b]{0.49\textwidth}
        \centering
        \includegraphics[width=\textwidth]{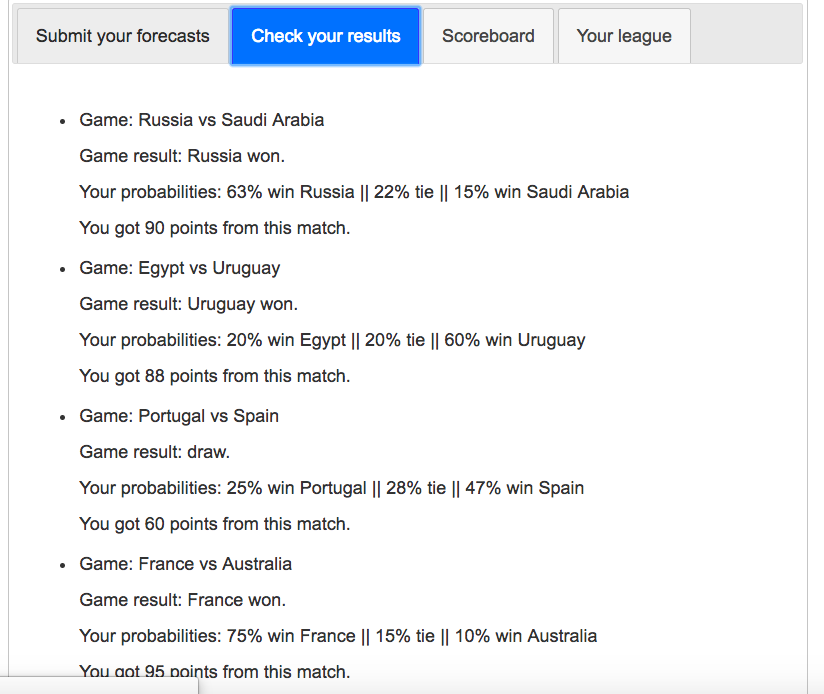}
        \caption{Checking results}
        \label{fig:3}
     \end{subfigure}
     \caption{Website entering probabilities and checking results interfaces}
\end{figure}

To save the informed probabilities, it was necessary to press the button ``Submit changes'', right below the text showing the points the forecast would score for each possible result of the matchup.
Some participants forgot to press this button, leaving the match with a blank forecast.
Figure \ref{fig:2} illustrates a forecast for the match Russia versus Saudi Arabia.

The second tab was ``Check your results'', where the bettor could find her probabilities already recorded by the system and, if the matches were already finished, the score of each forecast (Figure \ref{fig:3}).

The ``Scoreboard'' tab brought the overall classification and the tab ``Your league'' showed the classification only of the members of the league the bettor was part of.
This tab also allowed the creation of new leagues and, for the user or participant with admin status, to manage the participation of other users asking to join the league.

At the bottom of the page there were five links.
The first one, ``Objectives'', brought the goals of the project.
Our main goal was to use a popular sporting event to motivate young students and math teachers, especially at pre-university level, to learn and discuss the main aspects of probability theory.
A secondary objective was to collect data to model them them with probabilistic forecasting methods.

The second link, ``History'', is a text about the origins of football betting contests, including the ideas of Bruno de Finetti about a probabilistic contest for football.
The link ``People \& contact'' listed the team members and their respective tasks.
The link ``Rules'' reported the contest rules, most importantly, how the forecasts would be scored after each match and other important points, such as:

\begin{itemize}
\item bettors could submit or change forecasts up to 30 minutes before the start of each match;

\item after the group phase, the matches are playoffs. If they ended tied after 90 minutes, they were considered a draw;

\item if a bettor did not submit a forecast to a match, her score for that match was zero;

\item forecasts could not be more precise than whole percentages, and points gained for each game were rounded to the nearest integer.
\end{itemize}

The scoring rule will be discussed more carefully in the next section.
The last link was the ``Log-out'', to leave the system, or ``Join or sign-in'' for those not logged in.
It was possible to log-in using a Google or Facebook account.
In the last link, ``Terms of use and privacy'', we display what users' information were collected and how we intended to use them in case of acceptance. The full terms can be consulted at \url{https://fifaexperts.com/games/terms/}. The framework of the website was developed using Django by Marco Inácio and the texts were written by Marcio Diniz. Before the WCT, the other authors helped advertising the website.

\section{The scoring rule}
\label{sec:score}

The proposed contest had to rank the participants according to their ability to make probabilistic forecasts about the outcome $\theta$ of a match, where $\theta$ is a random variable taking values in $\Theta = \{\theta_1, \theta_2, \theta_3\}$, with $\theta_1$ standing for a victory of the first team, $\theta_2$ a victory of the second team and $\theta_3$ a draw. The probabilistic forecast for $\theta$ is represented by the vector $P = (P_1,P_2,P_3)$ of probabilities for each possible outcome $\theta_1, \theta_2$ and $\theta_3$. In order to rank the different probabilistic forecasts we use a scoring rule, which is a number that quantifies how ``far'' the declared forecast is from the match outcome. Even though this quantification can be done with different metrics \cite{machete2013}, the most natural way is by considering the Brier score \cite{Brier1950} defined as the squared Euclidean distance between the forecast and the outcome. For a review of scoring rules see (\cite{winkler1996, gneiting2007}).

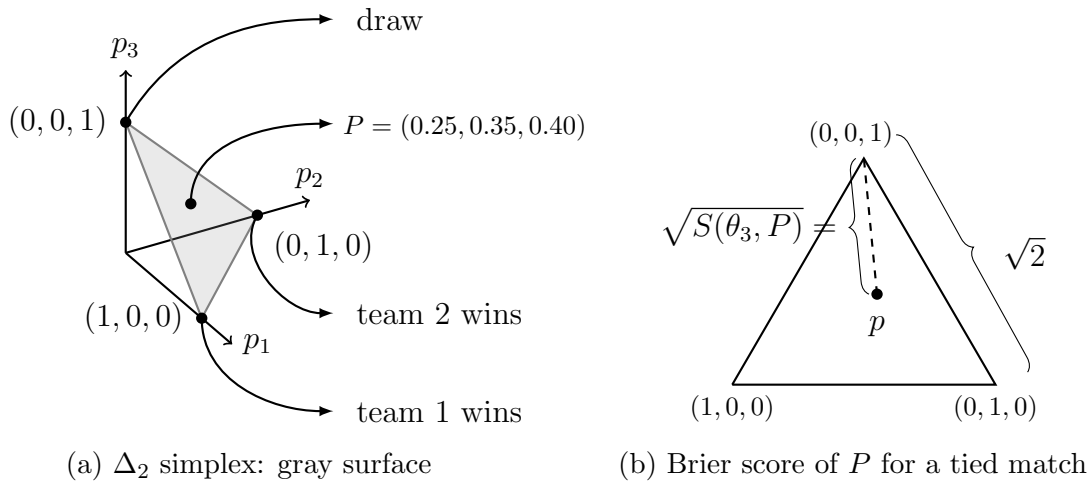
\begin{figure}[H]
		\centering
		\begin{subfigure}[t]{0.45\linewidth}
			\centering
			
\begin{tikzpicture}[scale=2,tdplot_main_coords,axis/.style={->},thick]
			
			\draw[axis] (0, 0, 0) -- (1.4, 0, 0) node [right] {$p_1$};
			\draw[axis] (0, 0, 0) -- (0, 1.4, 0) node [above] {$p_2$};
			\draw[axis] (0, 0, 0) -- (0, 0, 1.4) node [above] {$p_3$};
			
			\coordinate  (d1) at (1,0,0){};
			\coordinate  (d2) at (0,1,0){};
			\coordinate  (d3) at (0,0,1){};
			
			\fill[gray!80,opacity=0.2] (d1) -- (d2) -- (d3)-- cycle;
			
			\draw[-, gray, thick] (0,0,1) -- (1,0,0);
			\draw[-, gray, thick] (0,0,1) -- (0,1,0);
			\draw[-, gray ,thick] (1,0,0) -- (0,1,0);
			
			\node[fill,circle,inner sep=1.5pt,label={left:$(1,0,0)$}] at (d1) {};
			\node[fill,circle,inner sep=1.5pt,label={south east:$(0,1,0)$}] at (d2) {};
			\node[fill,circle,inner sep=1.5pt,label={left:$(0,0,1)$}] at (d3) {};
			
			\draw[-latex,thick](d3) to [out=60,in=180] (1,1,2);
			
			\node[label={right:draw}] at (1,1,2) {};
			
			\draw[-latex,thick](d1) to [out=-90,in=180] (1,1,-1);
			
			\node[label={right:team 1 wins}] at (1,1,-1) {};
			
			\draw[-latex,thick](d2) to [out=-120,in=180] (1,1,-.25);
			
			\node[label={right:team 2 wins}] at (1,1,-.25) {};
			
			\node[fill, black, circle,inner sep=1.5pt] at (0.25,0.35,0.40) {};
			
			\draw[-latex,thick](0.25,0.35,0.40) to [out=90,in=180] (1,1,1.2);
			
			\node[label={right:\footnotesize{$P=(0.25,0.35,0.40)$}}] at (1,.9,1.2) {};
			
			\end{tikzpicture}

\caption{$\Delta_2$ simplex: gray surface}\label{fig:3Dsimplex}
		\end{subfigure}	
		\hspace*{\fill}
				\begin{subfigure}[t]{0.45\linewidth} 
					\centering
					
					\begin{tikzpicture}[scale=3]
					\draw [thick](0,0) -- (1.1547,0) -- (0.57735,1)-- (0,0);
					
					\node[fill, black, circle,inner sep=1.5pt] at (0.63509,0.40) {};
					\node[label={below:$p$}] at (0.63509,0.40) {};
					\draw[decorate,decoration={brace,amplitude=5pt,mirror},xshift=4pt,yshift=3pt] (1.1547,-0.05) -- (0.57735,1) node[black,midway,right,xshift=0.4cm] {$\sqrt{2}$};

					\draw[dashed,thick] (0.63509,0.40) -- (0.57735,1);
					
					\draw[decorate,decoration={brace,amplitude=5pt},xshift=-1pt,yshift=0pt] (0.63509,0.40) -- (0.57735,1) node[black,midway,xshift=-1.5cm] {$\sqrt{S(\theta_3,P)}=$};

					
					\node[label={below:\footnotesize{$(1,0,0)$}}] at (0,0.05) {};
					\node[label={below:\footnotesize{$(0,1,0)$}}] at (1.1547,0.05) {};
					\node[label={above:\footnotesize{$(0,0,1)$}}] at (0.52,0.95) {};

					\end{tikzpicture}

					\caption{Brier score of $P$ for a tied match}
					\label{fig:2Dsimplex}
				\end{subfigure}
		\vspace*{\fill}
		\caption{$\Delta_2$ simplex viewed on three \textbf{(a)} and two \textbf{(b)} dimensions}
		\label{fig:norm_stand}
	\end{figure}

To formally define the Brier score, observe that the forecast $P=(P_1,P_2,P_3)$ lies in the 2-simplex, i.e.,  $P \in \Delta_2=\{(p_1,p_2,p_3):p_1+p_2+p_3=1, \ p_1, p_2, p_3 \geq 0\}$. Then, the Brier score $S(\theta,P)$ is given by

\begin{align*}
S(\theta,P) &= \sum_{i=1}^3\mathbb{I}(\theta=\theta_i)(1-P_i)^2+\sum_{i=1}^3\mathbb{I}(\theta\neq \theta_i)P^2_i 
\end{align*}
where $\mathbb{I}$ is the indicator function, i.e. it equals $1$ if the argument of the function is verified and zero otherwise.
Figure \ref{fig:3Dsimplex} illustrates the forecast $P=(0.25,0.35,0.40)$ represented on the 2-simplex.





When $P=(1,0,0)$ the forecaster puts all the probability at one vertex of the simplex, believing that $\theta=\theta_1$ for sure.
In this case, if team $1$ in fact wins the match, the Brier score is zero, its minimum possible value.
However, if the match is won by team two ($\theta=\theta_2)$ or ends tied ($\theta=\theta_3)$, the Brier score is $2$, its maximum possible value.

Therefore, for the Brier score, the smaller the scores, the better are the forecasts, which is somewhat counterintuitive for a score.
For this reason, we adopted a linear function of the Brier score as the scoring rule for our constest, namely
\begin{align}
\label{eq:our_score}
    S^*(\theta,P)=100-50\cdot S(\theta,P)
\end{align}
that is bounded between zero (worst score) and 100 (best score).
Since the World Cup had 64 matches, the perfect score would be 6400 points.

An important feature of the Brier score and our $S^*(\theta,P)$ is that they are {\it proper} \cite{dawid2014}, meaning that they lead the forecaster to inform her true probabilities to maximize the expected value of the score.

\section{Maximin strategy}
\label{sec:maximin}

1. The naive strategy $P^* = (1/3,1/3,1/3)$ is the non-randomized maxi-min strategy.

\

{\bf Proof}:

A strategy $P^*$ is maximin when
\begin{equation}
\inf_{\theta\in \Theta}[S^*(\theta, P^*)] \geq \inf_{\theta\in \Theta}[S^*(\theta, P)] \label{eq:maxmin1}
\end{equation}
for all $P \in \Delta_2$, that is, $P^*$ maximizes the minimum score given by $S^*(\theta, P) = 100 - 50 \cdot S(\theta,P)$. 
Notice that, $P^*$ satisfies \eqref{eq:maxmin1} if, and only if,
\begin{equation}
\sup_{\theta \in \Theta} S(\theta, P) \geq \sup_{\theta \in \Theta} S(\theta, P^*) \label{eq:maxmin2}
\end{equation}
for all $P \in \Delta_2$. 

Thus, the naive strategy $P^* = (1/3, 1/3, 1/3)$ is maximin if
\begin{equation}
\sup_{\theta \in \Theta} S(\theta, P) \geq \sup_{\theta \in \Theta} S(\theta, P^*) = \dfrac{2}{3} \label{eq:maxmin3}
\end{equation}
for all $P \in \Delta_2$. Observe that proving \eqref{eq:maxmin3} is equivalent to prove that for any $P = (P_1, P_2, P_3) \in \Delta_2$ we have
\begin{equation}
S(\theta_i, P) \geq \dfrac{2}{3} \text{ \ for some } i = 1, 2, 3. \label{eq:maxmin4}
\end{equation}

Associating the possible outcomes $\theta_1, \theta_2, \theta_3$ respectively to the points $O_1 = (1, 0, 0), \ O_2 = (0, 1, 0), \ O_3 = (0, 0, 1) \in \Delta_2$, we see that $S(\theta_i, P)$ is the squared Euclidean distance between the points $O_i$ and $P$, denoted by $\| \vv{P O_i} \|^2$. Then, in order to prove \eqref{eq:maxmin4}, we shall prove that $\| \vv{P O_i} \|^2 \geq 2/3$ for at least one $i = 1, 2, 3$.

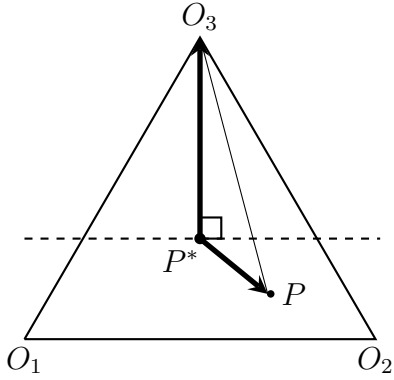
\begin{figure}[H]
    \centering
              
          \begin{tikzpicture}[scale=4]
          \draw [thick](0,0) -- (1.1547,0) -- (0.57735,1)-- (0,0);
          

          \coordinate (P) at (0.57735,1/3) {};
          \coordinate (P1) at (0.57735,0.5) {};
          \coordinate (P2) at (0.62,1/3) {};
          \tkzMarkRightAngle[thick,size=.07](P1,P,P2);

          \node[label={below:$O_1$}] at (0,0.05) {};
          \node[label={below:$O_2$}] at (1.1547,0.05) {};
          \node[label={above:$O_3$}] at (0.57735,0.95) {};
          
          \draw[dashed,thick] (0,1/3) -- (1.17,1/3);
          \node[fill, black, circle,inner sep=1.5pt] at (0.57735,1/3) {};
          \node[label={left:$P^*$}] at (0.65,0.26) {};
          \node[fill,black, circle,inner sep=1.0pt] at (0.81,0.15) {};
          \node[label={right:$P$}] at (0.77,0.15) {};
          \draw[line width=2pt,-stealth] (0.57735,1/3) -- (0.8,0.15);
          \draw[line width=2pt,-stealth] (0.57735,1/3) -- (0.57735,1);
          \draw (0.57735,1) -- (0.8,0.15);

          \end{tikzpicture}
          \caption{Illustration of vectors  $\protect\vv{P^* O_3}$, $\protect\vv{P O_3}$ and $\protect\vv{P^* P}$} \label{fig:simplex}
\end{figure}

Since $P_1 + P_2 + P_3 = 1$, then $P_i \leq 1/3$ for some $i$. For instance, we can assume that $P_3 \leq 1/3$. Then, it is possible to show that
\begin{equation}
\langle \vv{P^*P}, \vv{P^* O_3} \rangle = P_3 - \dfrac{1}{3} \leq 0, \label{eq:dotprod}
\end{equation}
where $\langle \cdot, \cdot \rangle$ is the dot product of the Euclidean space. The inequality \eqref{eq:dotprod} is equivalent to say that the angle between the vectors $\vv{P^* P}$ and $\vv{P^* O_3}$ is greater or equal to $\pi/2$. See Figure \ref{fig:simplex} for an illustration of the mentioned vectors on the simplex. 

From the polarization identity (law of cosines), it follows that
\begin{equation*}
\|\vv{P O_3} \|^2 = \| \vv{P^* P} \|^2 + \| \vv{P^* O_3} \|^2 - 2 \langle\vv{P^* P}, \vv{P^* O_3}\rangle \geq \|\vv{P^* O_3} \|^2,
\end{equation*}
where the last inequality follows from \eqref{eq:dotprod}. 
Then, 
$$
S(\theta_3, P) = \|\vv{P O_3} \|^2 \geq \|\vv{P^* O_3} \|^2 = S(\theta_3, P^*) = \dfrac{2}{3},
$$
which proves \eqref{eq:maxmin4}. Therefore, $P^* = (1/3,1/3,1/3)$ is a maximin strategy.

\begin{figure}[!h]
    \centering
    \includegraphics[scale=0.7]{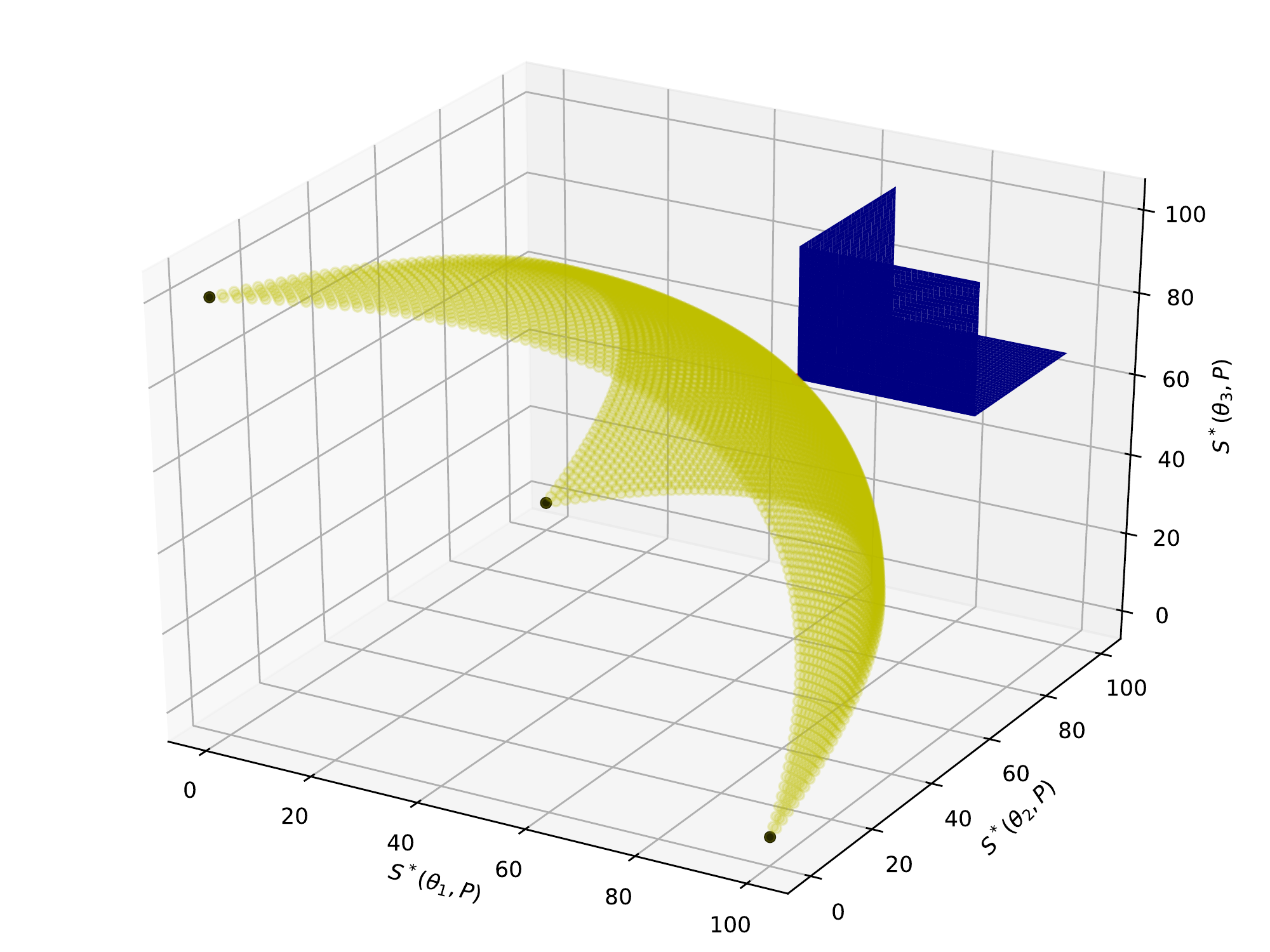}
    \caption{Score surface of each pure strategy $P$}
    \label{fig:minimax}
\end{figure}

Figure \ref{fig:minimax} displays the surface of scores $\big(S^*(\theta_1, P), \linebreak S^*(\theta_2, P), S^*(\theta_3, P)\big)$ with $P$ varying on the simplex $\Delta_2$ (yellow surface) and the level surface of  the score vectors such that $\min\{S^*(\theta_1, P), \linebreak S^*(\theta_2, P), S^*(\theta_3, P)\} = 200/3$ (blue surface), where the $200/3$ is the minimum score for the naive strategy $P^*$.
The intersection point of the two surfaces indicates $P^*$. 

\

\noindent
2. Expected scores of the mixed strategies: (i) Dirichlet(1,1,1); (ii) uniform on vertices; (iii) uniform of edges.

First we rewrite the expression for $S(\theta,P)$.

\begin{align*}
S(\theta,P)& =\sum_{i=1}^3\mathbb{I}(\theta=\theta_i)(1-P_i)^2+\sum_{i=1}^3\mathbb{I}(\theta\neq \theta_i)P^2_i\\
 &=1+\sum_{i=1}^3P_i^2-2\sum_{i=1}^3P_i ~ \mathbb{I}(\theta=\theta_i)
\end{align*}

Then, for $\theta\in\Theta=\{\theta_1,\theta_2,\theta_3\}$:

\[E[S(\theta,P)]=1+\sum_{i=1}^3E[P_i^2]-2\sum_{i=1}^3E[P_i] \cdot \mathbb{I}(\theta=\theta_i)\]

\

(i) Dirichlet (1,1,1)

\[E[S(\theta_1,P)]=E[S(\theta_2,P)]=E[S(\theta_3,P)]=5/6 \Longrightarrow E[S(\theta,P)]=5/6,\]

\noindent
and therefore

\[E[S^*(\theta,P)]=100-50\cdot E[S(\theta,P)]=58.33.\]

\

(ii) Mixed strategy: each vertex with probability $1/3$

\[E[S(\theta_1,P)]=E[S(\theta_2,P)]=E[S(\theta_3,P)]=4/3 \Longrightarrow E[S(\theta,P)]=4/3,\]

\noindent
and therefore

\[E[S^*(\theta,P)]=100-50\cdot E[S(\theta,P)]=33.33.\]

In fact, this strategy is the randomized maximin strategy, corresponding to a randomization of the three points highlighted on the utility surface displayed by Figure \ref{fig:minimax}.

\

(iii) Uniform on the edges

\begin{align*}
E[S(\theta_1,P)]=E[S(\theta_2,P)]=E[S(\theta_3,P)]=&\frac{1}{3}\int_0^1(0-u)^2+(1-(1-u))^2\,du\\ &+ \frac{1}{3}\int_0^1(0-(1-u))^2+(1-u)^2\,du\\ &+ \frac{1}{3}\int_0^11+(0-u)^2+(0-(1-u))^2\,du\\\ \Longrightarrow E[S(\theta,P)]=1,&
\end{align*}
\noindent
and therefore

\[E[S^*(\theta,P)]=100-50\cdot E[S(\theta,P)]=50.\]

\

\newpage

\newgeometry{top=40mm, bottom=30mm}

\section{2018 WTC matches}
\label{sec:extra_tables}

\begin{table}[!htb]
\vspace{-1.5em}
\centering
\footnotesize
\caption{Group phase matchups}
\begin{tabular}{crclc}
\hline
Match number 	& & Final result & & Forecasts\\
\hline
1	&	Russia	&	5 $\times$ 0	&	Saudi Arabia	&	363	\\
2	&	Egypt	&	0 $\times$ 1	&	Uruguay	&	358	\\
3	&	Morroco	&	0 $\times$ 1	&	Iran	&	347	\\
4	&	Portugal	&	3 $\times$ 3	&	Spain	&	354	\\
5	&	France	&	2 $\times$ 1	&	Australia	&	348	\\
6	&	Argentina	&	1 $\times$ 1	&	Iceland	&	342	\\
7	&	Peru	&	0 $\times$ 1	&	Denmark	&	332	\\
8	&	Croatia	&	2 $\times$ 0	&	Nigeria	&	340	\\
9	&	Costa Rica	&	0 $\times$ 1	&	Serbia	&	330	\\
10	&	Germany	&	0 $\times$ 1	&	Mexico	&	342	\\
11	&	Brazil	&	1 $\times$ 1	&	Switzerland	&	350	\\
12	&	Sweden	&	1 $\times$ 0	&	South Korea	&	333	\\
13	&	Belgium	&	3 $\times$ 0	&	Panama	&	341	\\
14	&	Tunisia	&	1 $\times$ 2	&	England	&	342	\\
15	&	Colombia	&	1 $\times$ 2	&	Japan	&	337	\\
16	&	Poland	&	1 $\times$ 2	&	Senegal	&	329	\\
17	&	Russia	&	3 $\times$ 1	&	Egypt	&	320	\\
18	&	Portugal	&	1 $\times$ 0	&	Morroco	&	318	\\
19	&	Uruguay	&	1 $\times$ 0	&	Saudi Arabia	&	311	\\
20	&	Iran	&	0 $\times$ 1	&	Spain	&	314	\\
21	&	Denmark	&	1 $\times$ 1	&	Australia	&	306	\\
22	&	France	&	1 $\times$ 0	&	Peru	&	318	\\
23	&	Argentina	&	0 $\times$ 3	&	Croatia	&	312	\\
24	&	Brazil	&	2 $\times$ 0	&	Costa Rica	&	334	\\
25	&	Nigeria	&	2 $\times$ 0	&	Iceland	&	305	\\
26	&	Serbia	&	1 $\times$ 2	&	Switzerland	&	299	\\
27	&	Belgium	&	5 $\times$ 2	&	Tunisia	&	302	\\
28	&	South Korea	&	1 $\times$ 2	&	Mexico	&	302	\\
29	&	Germany	&	2 $\times$ 1	&	Sweden	&	308	\\
30	&	England	&	6 $\times$ 1	&	Panama	&	304	\\
31	&	Japan	&	2 $\times$ 2	&	Senegal	&	292	\\
32	&	Poland	&	0 $\times$ 3	&	Colombia	&	295	\\
33	&	Uruguay	&	3 $\times$ 0	&	Russia	&	290	\\
34	&	Saudi Arabia	&	2 $\times$ 1	&	Egypt	&	287	\\
35	&	Iran	&	1 $\times$ 1	&	Portugal	&	301	\\
36	&	Spain	&	2 $\times$ 2	&	Morroco	&	300	\\
37	&	Denmark	&	0 $\times$ 0	&	France	&	295	\\
38	&	Australia	&	0 $\times$ 2	&	Peru	&	286	\\
39	&	Iceland	&	1 $\times$ 2	&	Croatia	&	287	\\
40	&	Mexico	&	0 $\times$ 3	&	Sweden	&	282	\\
41	&	South Korea	&	2 $\times$ 0	&	Germany	&	289	\\
42	&	Serbia	&	0 $\times$ 2	&	Brazil	&	296	\\
43	&	Switzerland	&	2 $\times$ 2	&	Costa Rica	&	282	\\
44	&	Japan	&	0 $\times$ 1	&	Poland	&	285	\\
45	&	Senegal	&	0 $\times$ 1	&	Colombia	&	287	\\
46	&	Panama	&	1 $\times$ 2	&	Tunisia	&	290	\\
47	&	England	&	0 $\times$ 1	&	Belgium	&	292	\\
48	&	Nigeria	&	1 $\times$ 2	&	Argentina	&	294	\\
\hline
\end{tabular}
\label{tab:group}
\end{table}

\begin{table}[!h]
\centering
\caption{Knock out matchups (Round of 16, quarter, semi and finals)}
\begin{tabular}{crclc}
\hline
Match number 	& & Final result & & Forecasts\\
\hline
49	&	France	&	4 $\times$ 3	&	Argentina	&	169	\\
50	&	Uruguay	&	2 $\times$ 1	&	Portugal	&	175	\\
51	&	Spain	&	1(3) $\times$ 1(4)$^{**}$	&	Russia	&	187	\\
52	&	Croatia	&	1(3) $\times$ 1(2)$^{**}$	&	Denmark	&	188	\\
53	&	Brazil	&	2 $\times$ 0	&	Mexico	&	187	\\
54	&	Sweden	&	1 $\times$ 0	&	Switzerland	&	185	\\
55	&	Belgium	&	3 $\times$ 2	&	Japan	&	184	\\
56	&	Colombia	&	1(3) $\times$ 1(4)$^{**}$	&	England	&	193	\\
57	&	Uruguay	&	0 $\times$ 2	&	France	&	158	\\
58	&	Russia	&	2(3) $\times$ 2(4)$^{**}$	&	Croatia	&	165	\\
59	&	Brazil	&	1 $\times$ 2	&	Belgium	&	165	\\
60	&	Sweden	&	0 $\times$ 2	&	England	&	151	\\
61	&	France	&	1 $\times$ 0	&	Belgium	&	131	\\
62	&	Croatia	&	2 $\times$ 1 $^*$	&	England	&	131	\\
63	&	Belgium	&	2 $\times$ 0	&	England	&	101	\\
64	&	France	&	4 $\times$ 2	&	Croatia	&	114	\\
\hline
\multicolumn{2}{l}{($*$) after extra time}\\
\multicolumn{3}{l}{($**$) after penalty kicks (score in parentheses)}
\end{tabular}
\label{tab:knock}
\end{table}

\end{document}